\documentclass[aip, jmp, amsmath,amssymb, reprint,floatfix,]{revtex4-1}

\usepackage{import}
\usepackage{graphicx}
\graphicspath{ {./Figures/} }
\usepackage{mathtools}
\usepackage{dcolumn}
\usepackage{bm}
\usepackage{braket}
\usepackage{enumitem}
\usepackage{float}
\usepackage[section]{placeins} 
\usepackage{booktabs}
\usepackage{array}
\usepackage{hyperref} 

\begin{document}

\title{Molecular Hessian matrices from a machine learning random forest regression algorithm}
\author{Giorgio Domenichini} 
\email{giorgio.domenichini@univie.ac.at}
\affiliation{Faculty of Physics,
University of Vienna,
Kolingasse 14-16,
1090 Vienna,
Austria}

\author{Christoph Dellago} 
\email{christoph.dellago@univie.ac.at}
\affiliation{Faculty of Physics,
University of Vienna,
Kolingasse 14-16,
1090 Vienna,
Austria}

\begin{abstract} In this article we present a machine learning model to obtain fast and accurate estimates of the molecular Hessian matrix. In this model, based on a random forest, the second derivatives of the energy with respect to redundant internal coordinates are learned individually. The internal coordinates together with their specific representation guarantee rotational and translational invariance. The model is trained on a subset of the QM7 data set, but is shown to be applicable to larger molecules picked from the QM9 data set. From the predicted Hessian it is also possible to obtain reasonable estimates of the vibrational frequencies, normal modes and zero point energies of the molecules.
\end{abstract}
\maketitle
\section{Introduction}

The molecular Hessian matrix, namely the matrix of second derivatives of the potential energy with respect to nuclear positions, has an important role in quantum chemistry. It describes the curvature of the Potential Energy Surface (PES), and the knowledge of it is important for geometry relaxation as well as for the determination of transition states\cite{schlegel2011geometry,reveles2004_DFT_geomopt}.
In the context of molecular geometry optimization, all quasi-Newton methods require the knowledge of an approximate Hessian matrix, as an inexpensive initial guess for the first step.
The knowledge of the Hessian matrix also helps to interpret infrared (IR) spectra of molecules: in the harmonic approximation, the vibrational frequencies and the vibrational normal modes are calculated from the eigenvalues and eigenvectors of the mass-weighted Hessian matrix \cite{pople_vibrations,Wilson1941_mwh,MCMURRY1959203_mwh}.

In general, molecular vibrations are motions of the whole molecule, but some normal modes are strongly localized on particular functional groups. For instance, bond stretching and some angle bending modes often involve only few atoms. Dihedral torsions, on the other hand, are more collective in character and are often mixed with other degrees of freedom in low-frequency collective mode, making it more difficult to identify them. The frequencies of localized vibrations depend on the atom species involved, on the bond order which connects the atoms, and to a smaller extent on the chemical environment. These frequency changes very little among different molecules, and tabulated values are available for them\cite{McMurry1952_correlation_IR}.


The \textit{ab initio} calculation of the Hessian is computationally expensive and therefore several methods to approximate the Hessian have been proposed.
For geometry optimization, the Hessian is often calculated using approximate quantum mechanical methods \cite{HEAD1986359,pople1988_optimization_HF_hessian, fischer_almlof1992_geomopt} or approaches based on force fields expressed in internal coordinates \cite{schlegel1984_hess_est,schlegel2005_hessian_update,LINDH1995423}.
To approximate the Hessian matrix in internal coordinates is particularly convenient, because they are rotationally and translationally invariant, and because they are less coupled than Cartesian coordinates (the Cartesian Hessian may have many non-diagonal elements comparable in magnitude with the diagonal ones).


An alternative way to compute an approximate Hessian is to use Machine Learning (ML), which is currently having a significant impact in many ares of 
computational chemistry.\cite{qml_ccs_anatole2018, qml_nutshell_rupp2015}
Many molecular properties can be predicted directly \cite{von2020exploring,qml_rupp_atomization, faber_qml_lower_DFT,christensen2019operators,weinreich2021machine,qml_properties,qml_optimization_hammer, qml_lemm2021energy, Keith2021QML} from ML models and recent works showed that it is possible to predict the shape of a PES using Redundant Internal Coordinates (RICs). \cite{mancini2020unsupervised,falbo2022integration} 
A common way to calculate the Hessian with ML methods is by differentiation of the potential energy expressed using kernel in Kernel Ridge Regression (KRR) models \cite{response_inccs,von2020thousands,heinen2021toward,Heinen2022_transition}, or neural networks. \cite{Westermayr2021_ml_exctstates} However, recent methods have been developed for the direct prediction of vibrational spectra \cite{Zhang2020_vibrational_nn,Han2022_MLvibrations_review}, and Hessian matrices, but not across chemical space . \cite{ML_DCVS_Gandolfi_Ceotto, ML_Ngas_Gandolfi_Ceotto}

 
In this paper we present an alternative ML approach to predict directly, element by element, the Hessian matrix in RICs, across chemical compound space.\cite{ccs} The central idea of the method is to learn the elements of the Hessian using a random forest of decision trees. Trained on Hessians computed quantum mechanically for subset of small molecules (up to 7 non-hydrogen atoms) extracted from the QM7 dataset\cite{QM7_blum,QM7_rupp}, the model succeeds in predicting vibrational frequencies and Zero point vibrational energies (ZPE) for some larger molecules, sampled from the QM9 \cite{QM9_Ruddigkeit,QM9_ramakrishnan2014quantum} dataset (molecules up to 9 non-hydrogen atoms).

The remainder of this article is organized as follows. In Section II we briefly summarize the definitions of RICs, and the transformations of the Hessian between coordinate systems. We also present a new set of coordinate-specific representations that can be used to train a Random Forest Regression (RFR) model for the diagonal elements of the Hessian matrix. In Section III we validate the prediction for the diagonal Hessian elements, as well as for some of the most important non-diagonal elements. Some conclusions are provided in Section IV.

\section{Methods}
\subsection{Redundant internal Coordinates} \label{sect:RICs}


The choice of internal coordinates is not unique, and many types of internal coordinates have been developed, such as the Z-matrix\cite{baker_z_matrix}, the normal coordinates \cite{sellers1978normal}, the natural internal coordinates developed by Pulay and coworkers \cite{pulay1979systematic,pulay1992geometry,fogarasi1992calculation}, the delocalized internal coordinates developed by Baker \textit{et al.} \cite{baker_delocalized}, and others.\cite{reveles2004_DFT_geomopt}

The redundant internal coordinates introduced by Schlegel \cite{schlegel1982_Optimization}, which we use in this work, describe the geometry of a molecule in terms of bonds, bond angles and dihedrals, which can be easily and uniquely defined.
The number of internal coordinates exceeds the number of degrees of freedom of the molecule. This redundancy, however, guarantees a complete description of molecular distortions.
In the context of geometry optimization, RICs are commonly used to build the initial guess for the Hessian matrix. \cite{schlegel1984_hess_est,Geom_opt_largemols_Schegel,schlegel2011geometry}

\subsubsection{Definition} \label{sect:RIC_defin}

The RICs used in this work are defined as follows \cite{Peng}:
\begin{enumerate} 
    \item A bond $I–J$ is established if the distance between two atoms $I$ and $J$ is shorter than 1.3 times the sum of their covalent radii. 
    \item An angle $I–J–K$ spanned by atoms $I$, $J$ and $K$ is established if atom $I$ and $J$ as well as $J$ and $K$ are bonded and the angle is larger than 45°.  
    \item A dihedral is established from two consecutive angles $I–J–K$ and $J-K-L$. If one of the two angles is 180°, i.e., three atoms lie on a line, they are used as a new base for a dihedral ($I–J=K=L-M$).
\end{enumerate}
Molecules with dihedrals defined using 5 atoms (\textit{e.g.} $ I–J=K=L-M$ ) will be excluded from the dataset of this article, because they generate inconsistencies in the machine learning representation. A separate treatment would be required for them. 
 
\subsubsection{Derivatives expression}
In this work, Hessians were calculated using DFT in Cartesian coordinates, then transformed to RICs to be used as training data for the ML algorithm.
The ML model predicts Hessians in RICs, which have to be transformed back to Cartesian coordinates, to calculate vibrational frequencies and zero point energies.
In this section we recall the Hessian transformations used here, which follow the formalism developed by Pulay, Fogarasi and Schlegel. \cite{fogarasi1992calculation,pulay1992geometry,Peng,schlegel_1998_vibr_analisis,allen1993_coord_transformations}

For a given set of Cartesian coordinates $x$ and RICs $q$, the Wilson $B$ matrix,\cite{wilson1980molecular, wilson1955molecular} is defined as follows: 
\begin{equation}
    B_{ij} :=\frac{\partial q_i}{\partial x_j}. 
\end{equation}
The transformation of the gradient from internal ($\mathbf{g}_q$) to Cartesian coordinates ($\mathbf{g}_x$) is then given by:
\begin{equation}
    \mathbf{g}_x=B^T \mathbf{g}_q.
\end{equation}
Since in general $B$ is not a square matrix, we define for $B$ a right pseudo-inverse matrix $B_{\rm inv}$, 
\begin{equation}
\begin{aligned}
    B_{\rm inv} &= B^T (B B^T)^{-1}, \\
    B_{\rm inv}^T &= (B B^T)^{-1} B.
\end{aligned}  
\end{equation}
The transformation of the gradient from Cartesian coordinates back to RICs is:
\begin{equation}
    \mathbf{g}_q=B_{\rm inv}^T \mathbf{g}_x. 
\end{equation}
Defining $B'$ as the derivative of $B$ with respect to the Cartesian coordinates, \
\begin{equation}
  B'_{ijk}=\partial^2 q_i /\partial x_j \partial x_k,
\end{equation}
the transformation of the Hessian matrix $H$ from RICs to Cartesian coordinates can be written as:
\begin{equation}
    H_x=B^T H_q B+(B')^T\mathbf{g}_q.
\end{equation} 
Accordingly, the back transformation from Cartesian coordinates to RICs reads
\begin{equation} \label{eq:hess_to_cartesian}
    H_q=B_{inv}^T (H_x-(B')^T\mathbf{g}_q)B_{\rm inv}.
\end{equation}
\subsection{Representations}   
 \label{sec:Representations}
It is a convenient choice to predict the Hessian in RICs element by element using a specific local representation for every matrix element of interest.
Since predicting all elements of the Hessian can be a challenging task, we will focus only on some of the most important terms while we set to zero the less relevant elements. 
The diagonal elements are the most important ones and they will be treated in detail, but we note our machine learning method can be extended also to non-diagonal terms (see Sec.\ref{sec:nondiag}).

Every diagonal element of the RIC Hessian corresponds to the second derivative of the energy with respect to an internal coordinate (bond, angle or dihedral).
For each internal coordinate we construct a specific representation describing it and its chemical environment. These representations contain lists of molecular parameters, such as nuclear charges, bond lengths, bond orders, bond angles, and dihedrals. 

Inside the representations, nuclear charges and bond lengths are expressed in atomic units (proton charges and Bohr), bond angles and dihedrals are expressed using the periodic form $1+\text{cos}(\alpha)$, which prevents a discontinuity at $\alpha =\pi$, similarly to previously developed symmetry functions. \cite{dellago_NN_polymorphic,Behler_symmfuns}
Bond orders were calculated using the formula from Ref. \cite{Mayer_bond_order_matrix}, which for a closed shell system has the form
\begin{equation}
  O_{IJ}=\sum_{\mu \in I, \nu \in J} (PS)_{\mu\nu} (PS)_{\nu\mu}
\end{equation}
where he indices $I$ and $J$ refer to atoms, $P$ is the monoelectronic density matrix and $S$ is the atomic orbital overlap matrix.
In order to improve precision, we imposed a model selection, prior to the actual regression, based on two criteria.
The first is a classification based on the type of atoms which define the coordinate, {\textit {e.g.}}, to predict the C-O bond component of the Hessian, we use a model trained only on other C-O bonds.
The second classification is whether the internal coordinate is part a ring system. Different machine learning models are associated with internal coordinates which have or have not two or more atoms belonging to the same ring system (we call them ring coordinates and acyclic coordinates, respectively). 

\subsubsection{Bond-bond elements}
To represent the bond between two atoms $I$ and $J$ correctly, it is important to create a rule to label $I$ and $J$ uniquely. The choice was to label $I$ and $J$ according to their:
\begin{enumerate} 
    \item Nuclear charge:  $Z_I\geq Z_J$.
    \item Hybridization: if $Z_I = Z_J$, $I$ must have an $s$ character higher than $J$.  
    \item Nuclear charges of the bonded atoms: if $I$ and $J$ have the same nuclear charge and hybridization, the atoms bonded to $I$ should have a higher nuclear charge than the ones bonded to $J$.
  \end{enumerate}

The representation is a 26-element vector, which collects the parameters of $I-J$ and the spatial arrangement of the atoms bonded to $I$ and $J$.
The first two elements of the representation are the bond length and order of $I-J$. 
After that, for every atom $I_n$ bonded to $I$ and different from $J$ the following parameters are added:
\begin{enumerate} 
    \item its nuclear charge,
    \item the bonding order with $I$,
    \item the distance from $I$, and
    \item the width of the angle $I_n-I-J$ with $I$ and $J$
\end{enumerate}
Then, the analogous parameters are added or every atom $J_n$ bonded to $J$. The atoms $I_n$ ($J_n$) are included in decreasing order of nuclear charge and bond order with $I$ ($J$). 
Since molecules in the QM7 dataset do not have atoms with more than 4 bonds, $n$ has values form 1 to 3; if $I$ or $J$ have less than 4 bonded atoms, null values are inserted to maintain the same representation size.
\begin{figure}[h]
    \centering
    \includegraphics[width=\linewidth]{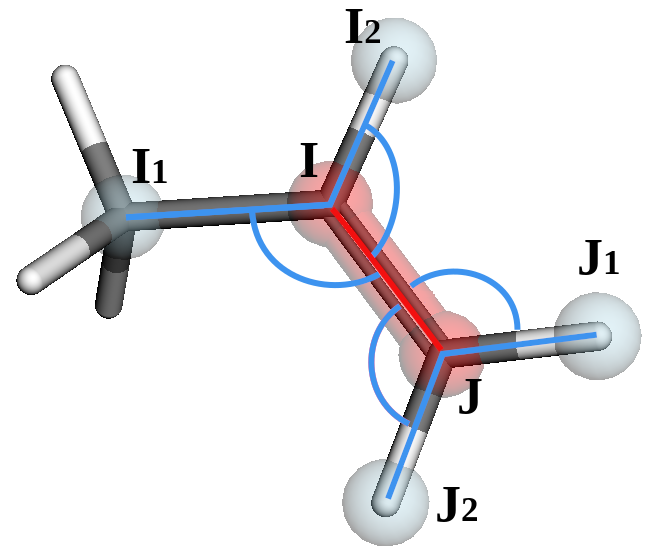}
    \includegraphics[width=\linewidth]{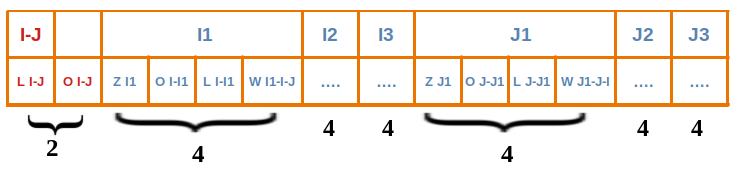}
    \caption{The representation for the $I-J$ bond, used to predict the bond-bond term in the Hessian, is built from listing bond lengths (L), bond orders (O), atom charges (Z) and angles (W), for $I$, $J$ and all atoms $I_n$ and $J_n$ bonded to $I$ and $J$}
    \label{fig:bond_repres}
\end{figure}


\subsubsection{Angle-angle elements}

In order to give a unique representation to the angle $I-J-K$, the atoms $I$ and $K$ need to be labelled uniquely according to the following criteria:
\begin{enumerate}
    \item Nuclear charge:  $Z_I\geq Z_K$
    \item Bond order with $J$: ${O}_{IJ}>{O}_{JK} $
    \item Hybridization:  $I$ must have an $s$ character higher than $K$ 
    \item Nuclear charges of the bonded atoms: the atoms bonded to $I$ should have a higher nuclear charge than the ones bonded to $K$
\end{enumerate}

Each angle representation consists of 46 elements. The first elements are the magnitude of the angle, the bond lengths $I-J$, $J-K$, and the bond orders $O_{IJ}$,$O_{JK}$,$O_{IK}$.
After these six parameters, for every atom $I_n$,$J_n$,$K_n$ bonded to $I$,$J$, or $K$, respectively, the several other quantities are listed. More specifically, for an atom $I_n$ bonded to atom $I$ (and different from $J$) the list include
\begin{enumerate}
    \item the nuclear charge of $I_n$
    \item the bond order of $I-I_n$
    \item the distance $I-I_n$
    \item the angle $J-I-I_n$ 
    \item the dihedral $K-J-I-I_n$ 
\end{enumerate}
For an atom $J_n$ bonded to $J5$ (different from $I$,$K$), the quantities are: the nuclear charge of $J_n$,the bond order of $J-J_n$,the distance $J-J_n$,the angle $K-J-J_n$ ,the dihedral $I-J-K-J_n$ .


Finally, for an atom $K_n$ bonded to $K$ (different from $J$), are added to the representation: the nuclear charge of $K_n$, the bond order of $K-K_n$, the distance $K-K_n$, the angle $J-K-K_n$, and the dihedral $I-J-K-K_n$. 
Since each atom has forms no more than four bonds, there are at most eight atoms bonded to $I$, $J$ or $K$ and in total the list includes at most 40 elements, which together with the previous 6 elements gives a representation of length 46. (Note, that if there are less then 8 atoms bonded to $I$, $J$, or $K$ the remaining elements are filled with zeros.) 

\begin{figure}[h]
    \centering
    \includegraphics[width=0.95\linewidth]{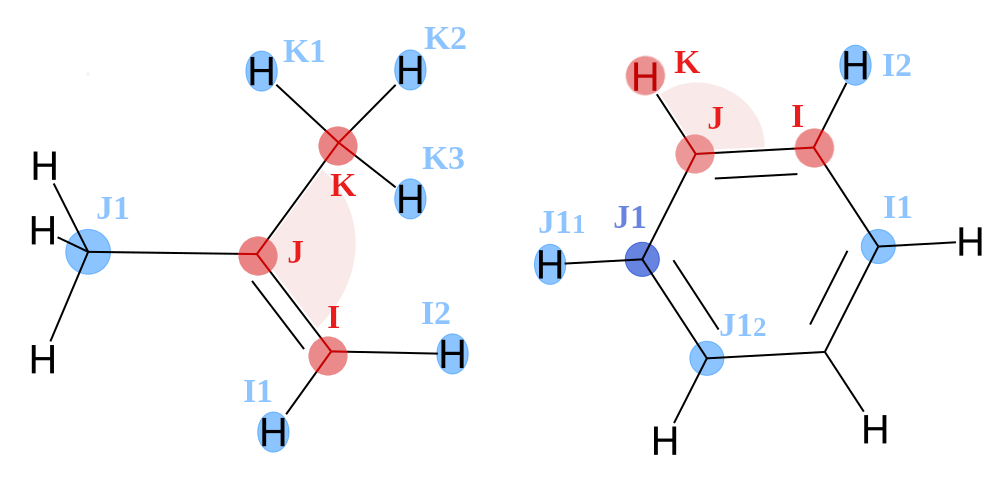}
    \caption{The representations for the angles $I-J-K$ is built listing the above mentioned parameters for the atoms $I_n$,$J_n$,$K_n$ (highligheted in light blue), bonded to $I$,$J$, or $K$. In the benzene molecule, where $K$ is a hydrogen, the $K_n$ are replaced in the representation by the $J1_n$ bonded to $J_1$.}
    \label{fig:angle_representation}
\end{figure}

If $K$ is a hydrogen atom, it does not have any bonded atoms ($K_n$) other than $J$ and the representation defined above would have many zeros. To compensate for this lack of descriptors, we include the five parameters of the lists above for the atoms $J1_n$ bonded to $J_1$ as shown in the second panel of Fig \ref{fig:angle_representation}. If both $K$ and $I$ are hydrogens, we include the five parameters of the lists above for the atoms $J1_n$ bonded to $J_1$, and also for the $J2_n$ bonded to $J_2$.

\subsubsection{Dihedral-dihedral elements}
The representation of a dihedral, used to predict the diagonal dihedral-dihedral terms of the Hessian, was built in a similar way to what was done for bonds and angles.
The label ordering of the dihedral $I-J-K-L$ was chosen such that $Z_J \ge Z_K$, and if $Z_J = Z_K$, than $Z_I \ge Z_L$. In case of a symmetrical atom arrangement, where $Z_J = Z_K$ and $Z_I = Z_L$, it is imposed that the bond orders $O_{IJ}>O_{KL}$.
The representation for a dihedral consists 65 elements.
The magnitude of the dihedral, the bond lengths and bond orders for the bonds $I-J$,$J-K$ and $K-L$, as well as the angles  $I-J-K$ and $J-K-L$ constitute the first 9 elements. 
Then other 6 elements describe how and if the dihedral is part of a ring system.
Finally, 50 more elements are given by all the up to 10 atoms bonded to $I,J,K$ or $L$, of which we list nuclear charges, the orders and lengths of the bonds to $I,J,K$ or $L$, the angles and the dihedrals formed with $I,J,K$ or $L$.

 For further clarification on the definition of the representations the reader is encouraged to consult the code and the dataset published on Zenodo\cite{domenichini_dataset_hessian}.
\subsection{Random Forest of decision trees}

The machine learning was performed using a random forest regression model\cite{RFR_Kang,RFR_Dutschmann,RFR_Breiman2001}, as it is implemented in SciKitLearn \cite{scikit-learn}.
Random forests have been used for the prediction of several molecular properties. \cite{RFR_MOLS_Faber,faber_qml_lower_DFT,RFR_SOLUB_Palmer,RFR_Ward_mols}

We found that in our case random forest regression performs better than neural networks and kernel ridge regression models. In fact, random forests allow more flexible representations and usually produce accurate results even for small training set sizes.

Training a random forest is also fast since the creation of every individual decision tree can be performed in parallel. In general, more decision trees in the forest lead to better (until the model saturates), but more expensive predictions.

SciKitLearn default hyper-parameters were used: those impose no restrictions on the sizes of the leaves, on the depths trees, or on the number of features chosen to perform a split; each forest contained 100 decision trees (estimators), trained with bootstrapping.

We chose then 100 decision trees as a compromise between precision and computing time. In fact, the Mean Absolute Error (MAE) for 100 decision trees is higher than the MAE for 1000 decision trees by approximately 1\% (see SI), but the predictions are ten times faster.

\subsection{Computational Details}
All quantum mechanical calculations were performed using PySCF \cite{pyscf_article}. DFT calculations were executed at the B3LYP\cite{becke1993,lee1988} level of theory, with the Dunning's correlation consistent cc-pVDZ atomic orbital basis set \cite{Dunning_1989}.
Geometry relaxation, RICs selection, and coordinate transformations were performed using a locally modified version of PyBerny \cite{hermann2020pyberny}. 
Machine learning was performed using SciKitLearn \cite{scikit-learn}.
All through our work were used Python\cite{van1995python,ipython} 3.8 scripts, together with Numpy\cite {harris2020numpy}, Scipy\cite{2020SciPy-NMeth} and Matplotlib\cite{hunter2007matplotlib} libraries.
Molecules were plotted using RdKit \cite{rdkit,rdkit, landrum2013rdkit} and Py3DMol \cite {py3dmol}, a Python interface to the Java-based 3Dmol\cite{rego20153dmol}. Molecular IUPAC names were generated using Stout\cite{stout}.

\section{Numerical Results}
\subsection{Cross validation of the Hessian's diagonal terms} \label{sec:CV}
The dataset for training and testing was built from the 6810 QM7 molecules, whose dihedrals are defined only by consecutive atoms (Section \ref{sect:RIC_defin}). The structures of this subset were relaxed and the Hessians were calculated at the B3LYP/cc-pVDZ level of theory. 
The molecules were then distributed randomly between training set and the test set. Since internal coordinates of the same molecule are in general not independent quantities, it is important that the train-test splits are performed on molecules and not on individual coordinates, so that coordinates of the same molecule either all belong to the training set, or they all belong to the test set.
The internal coordinates were classified according to their chemical composition and to the inclusion in a ring system (as explained in section \ref{sec:Representations}).
For each of these groups, different machine-learning models were trained independently.

\begin{figure}[h!]
    \centering
    \includegraphics[width=\linewidth]{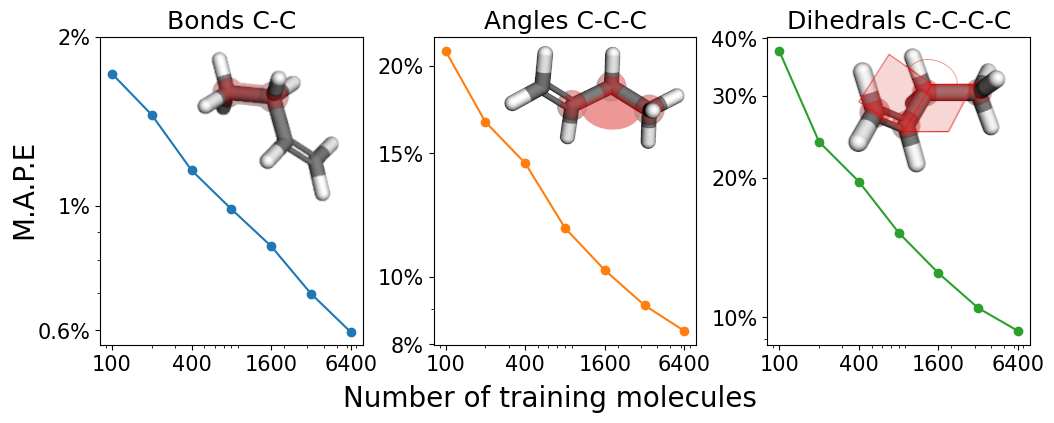}
    \caption{Learning curves for bonds, angles and dihedrals between carbon atoms not included in a ring. The data come from the average of ten train-test splits of our dataset for various training set sizes and a fixed test set size of 400 molecules.}
    \label{fig:learning_curves}
\end{figure}

The learning can be visualized as the decrease of the prediction error with increasing training set size. 

Focusing on some of the most frequently occurring RICs in the training set (acyclic C-C bonds, C-C-C angles, and C-C-C-C dihedrals), in Figure \ref{fig:learning_curves} we show the Mean Absolute Percentage Error (MAPE) for the predictions of the Hessian elements as a function of the number of molecules in the training set.

In an ideal machine learning model the learning curves \cite{learningcurves_bing,learningcurves_krr_muller} should exhibit, in a log-log plot, a linearly decaying trend. In our cases, this linear trend is approximated very well for bonds; for angles and dihedrals there is a small deviation, but still learning is achieved throughout the chosen training set. 

\begin{figure}[h!]
    \centering
    \includegraphics[width=\linewidth]{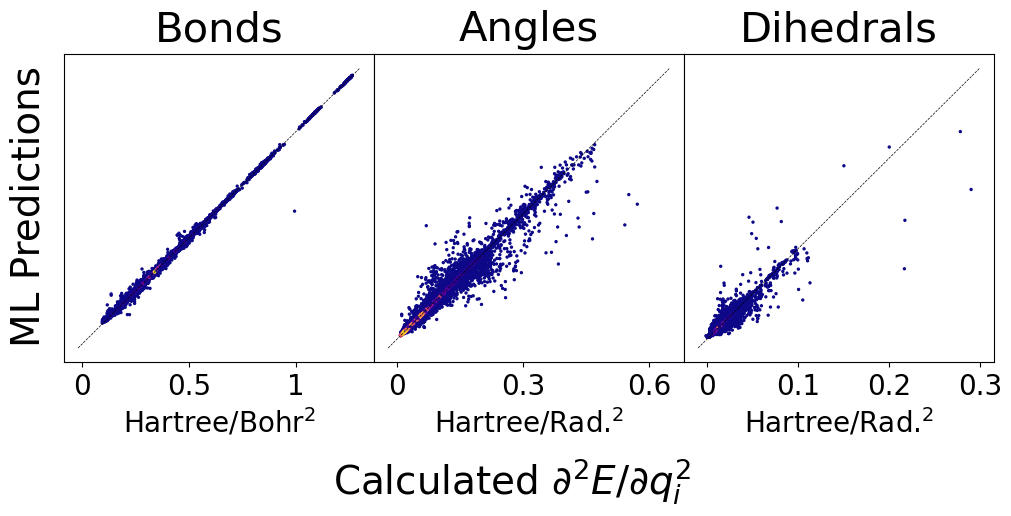}
    \caption{
    ML predictions vs. the calculated values of all diagonal Hessian elements, plotted for one particular train-test split. }
  \label{fig:Diag_Scatter}
\end{figure}

Diagonal elements of the Hessian predicted by the machine learning models for all bonds, angle and dihedrals are compared to the reference data in Figure \ref{fig:Diag_Scatter}.
As can be inferred from the figure, the accuracy is higher for bonds than for angles and and particularly for dihedrals. 
For bond derivatives the only outlier is a single C-C bond inside a conjugated atom chain. The reason is that the machine learning model treats this bond as a single bond while the true value of this Hessian element is closer to that of a double bond. 

\begin{figure}[h!]
    \centering
    \includegraphics[width=\linewidth]{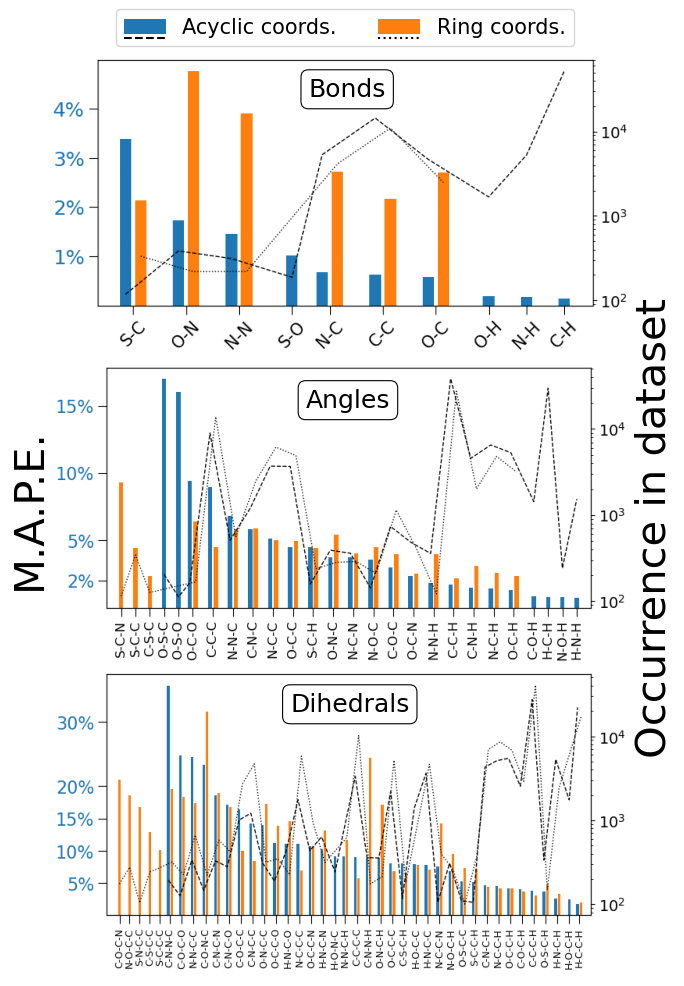}
    \caption{MAPEs for Hessian elements belonging to bonds, angle and dihedrals for different combination of elements. Each MAPE is obtained as average over 10 different train-test splits (fixed 3:1 size ratio). Only MAPEs for coordinates  that appear at least 100 times in the dataset are shown.}
    \label{fig:Bar_mape}
\end{figure}

The performance of our method was estimated for all coordinates and all atomic element combinations within our QM7 dataset. The MAPEs resulting from ten train-test splits are summarized in Figure \ref{fig:Bar_mape}. 

The prediction error is around 2\% for bond-bond, around 5\% for angle-angle, and around 10\% for dihedral-dihedral terms.
Bond-bond Hessian elements can be predicted more accurately, probably because bond stretching is a more localized motion than angle scissoring or dihedral torsion, which can be associated with wider molecular motions and may depend more strongly on the non-bonded atomic surroundings.  
For similar reasons, the most accurate predictions are those for bonds between hydrogens and a heavier element (C, N, O), the MAPEs of which are below 0.2\%. 

Hessian elements related to C-N, C-C, and C-O bonds are also well predicted (with an error of about 0.5\% if not inside a ring), because they occur frequently in the training set.

On the contrary, the biggest errors were made for S-C, O-N, and N-N bonds, because their occurrence in the training set is limited.

Also for angles and dihedrals, those including hydrogen atoms have the lowest prediction errors. In fact, the rotation of a hydrogen atom is usually not correlated with other molecular motions and thus it is more consistently repeated in chemical space. The highest errors were obtained for dihedrals and angles which have a limited number of occurrences in the dataset, mostly involving S, O, and N angles.


\subsection{Non-diagonal terms}
\label{sec:nondiag}
\begin{figure}[ht]
    \centering
    \includegraphics[width=\linewidth]{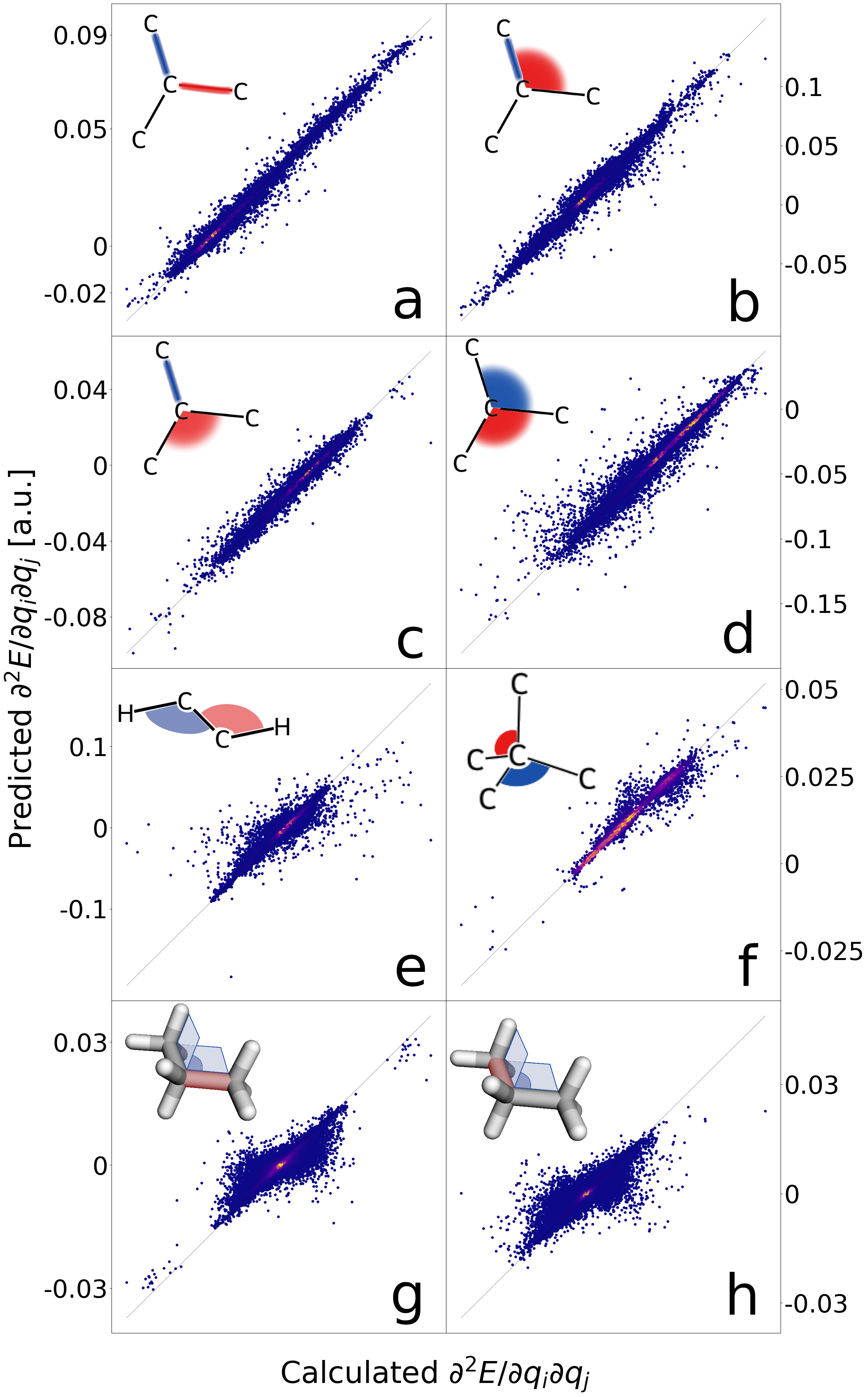}
    \caption{Predicted vs. calculated non-diagonal terms of the Hessian for different combinations of internal variables: a) consecutive bond-bond, b) included bond-angle, c) adjacent bond-angle, d) adjacent angle-angle, e) consecutive angle-angle, f) opposite angle-angle, g) external bond-dihedral, and h) internal bond-dihedral. The scatter plots shown were obtained for a single train test split of our dataset.}
    \label{fig:Nondiags}
\end{figure}

While the non-diagonal elements of the Hessian are usually small and often assumed zero in many approximations \cite{schlegel1984_hess_est}, some of them can be successfully predicted using the machine learning framework presented in this article. We applied machine learning to some of the most important non-diagonal elements, assuming that the geometrical proximity of the internal coordinates is correlated with the magnitude of the Hessian term.
In particular, we predicted the following terms of the Hessian: consecutive (sharing one atom) bond-bond terms; included (the bond is a side of the angle) and adjacent (sharing one vertex)  bond-angle terms; adjacent (sharing one side and the vertex), consecutive (sharing one side), and opposite (sharing the vertex) angle-angle terms; and the mixed terms between a dihedral angle and the bonds (internal and external) between the atoms which defining the dihedral. The combinations of internal coordinates, for which we predicted the non-diagonal Hessian elements, are schematically shown in the various panels of Fig. \ref{fig:Nondiags}.
For each type of these non-diagonal terms we constructed a specific representation, built from the enumeration of the type and the positions of the atoms surrounding the involved internal coordinates, similar to what was done for the diagonal terms in Sec.\ref{sec:Representations}. As for the diagonal terms of the Hessian, coordinates pairs are now classified according to the chemical elements which constitute them, so that different chemical elements are predicted with different machine learning models. We refer the reader to the code published on the Zenodo repository for a detailed definition of the representation \cite{domenichini_dataset_hessian}.

\begin{table}[h]
    \centering
    \begin{tabular}{l|c}
    Coordinate Pair & MAE (  10$^{-3}$ a.u.) \\
    \hline
     Bond-bond consecutive & 0.54  \\
     Bond-angle included & 0.86  \\
     Bond-angle  adjacent & 0.47 \\
     Angle-angle adjacent & 0.99  \\
     Angle-angle consecutive & 1.66 \\
     Angle-angle opposite & 0.89  \\
     Bond-dihedral external & 0.69  \\
     Bond-dihedral internal & 0.87  \\
    \end{tabular}
    \caption{Mean absolute error for the prediction of the non-diagonal Hessian terms, averaged over 10 train-test splits of the dataset.}
    \label{tab:nondiag_MAE}
\end{table}

The scatter plots shown in Figure \ref{fig:Nondiags} are for one train test split of our data. In the figure, data obtained for every possible combination of atomic species, are shown together. Values are expressed in atomic units (a.u.), which correspond to Hartree/Bohr$^2$, Hartree/(Bohr*Radians), Hartree/Radians$^2$, depending on the type of coordinate pairs involved.
 
As can be inferred from Figure \ref{fig:Nondiags}, the correlation between true (calculated) and predicted values is particularly good for panels a, b and c, which are related to bond-bond and bond-angle terms. The errors are largest in the prediction of consecutive angle-angle terms (panel e), and bond-dihedral elements (panels g and h). All of these non-diagonal elements are small in magnitude and can be zero. 
For this reason, instead of the percentage error, we report in Table \ref{tab:nondiag_MAE} the mean absolute error (MAE), averaged over 10 train test splits. This error is on the order of $10^{-3}$ a.u., which is small compared to the range $\pm 0.1$ a.u. of possible values shown in Fig. \ref{fig:Nondiags}.
\subsection{Prediction of QM9 Hessians}
\begin{figure}[h]
    \centering
    \includegraphics[width=.5\textwidth]{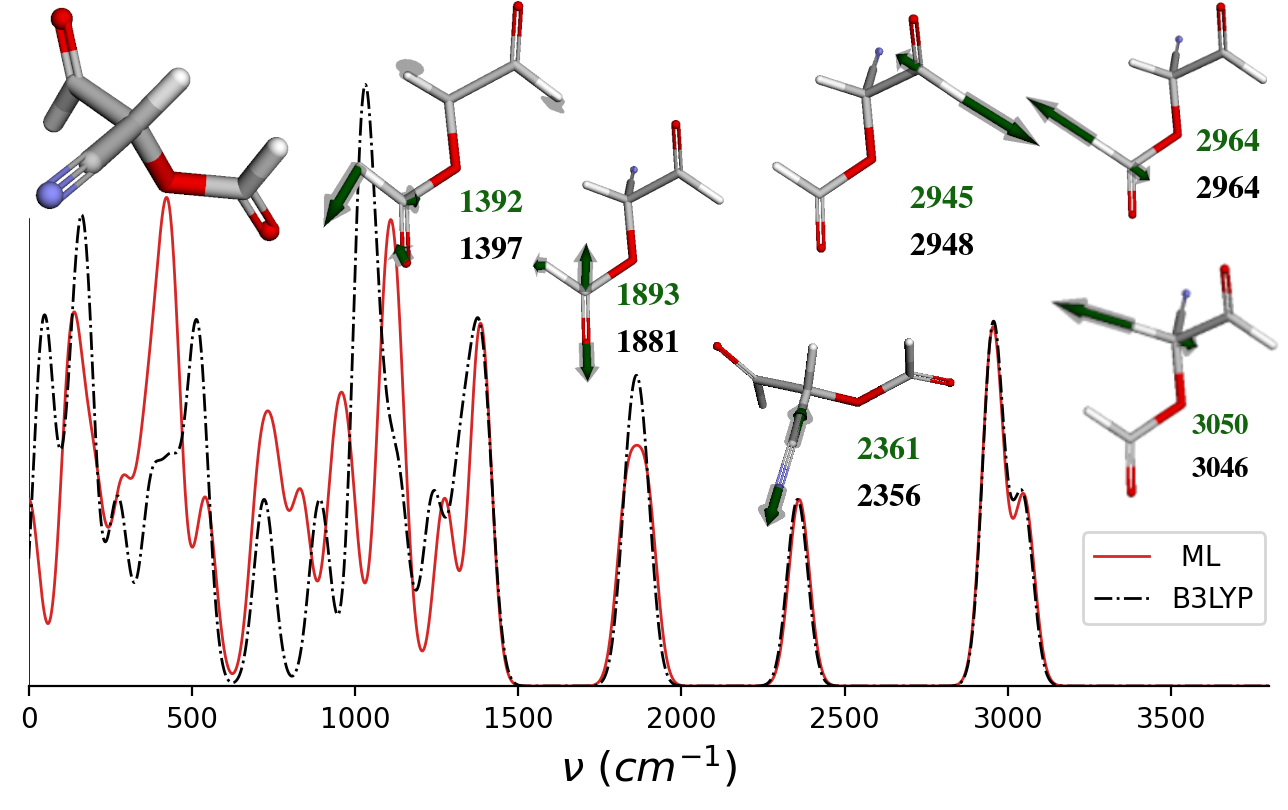}
    \caption[width=\linewidth]{Vibrational frequencies of molecule Nr. 10,000 of the QM9 dataset predicted by ML (solid red line) and calculated with B3LYP-DFT (dashed black line). The frequencies were convoluted with a Gaussian with fixed  height and standard deviation (equal to 32cm$^-1$).    
    In the figure are also depicted some selected normal modes of the ML predicted (green arrows) and of the calculated (grey arrows) frequencies. }
    \label{fig:QM9_vibrfreq}
\end{figure}

Given the locality of our representation, it is possible to predict the Hessian matrix for molecules of any given size. While we trained the machine learning model on the QM7 data set, we will now test it by predicting the Hessian for molecules of the QM9 dataset. As an example, we picked the molecules of QM9 whose index is a multiple of 10,000, from molecule 10,000 to 130,000. These 13 molecules possess 8 or 9 heavy atoms, so they are bigger than any molecule in the training set.

At first, we optimized the molecular geometries using the B3LYP/cc-pVDZ level of theory, and  calculated the Hessians explicitly, in order to compared them with the ML predictions.
For the ML Hessian expressed in RICs, we calculated all the diagonal elements as well as the non-diagonal elements listed in section \ref{sec:nondiag}.
The transformation from internal ($H_q$) to Cartesian ($H_x$) coordinates was performed according to Equation \ref{eq:hess_to_cartesian}.
After the transformation to Cartesian coordinates, it is possible to calculate the expected harmonic vibrational frequencies, as the eigenvalues of the mass-weighted Hessian matrix while the corresponding eigenvectors correspond to the vibrational normal modes\cite{MCMURRY1959203_mwh,Wilson1941_mwh,li2002partial_mwh} (Eq.\ref{eq:red_mass_H}). 
\begin{equation} \label{eq:red_mass_H}
  H^\mathrm{MW}_{x_Ix_J} := \frac{H_{x_Ix_J}}{\sqrt{m_I {m_J}}}
\end{equation}
Here, $x_I$ is a Cartesian coordinate of atom $I$ and $m_I$ is its atomic mass.

While the simulation of an IR spectrum should include the explicit calculation absorption intensities and peak broadenings \cite{herman1955influence,segal1967calculation,IRabsorb_vibrfreq_Ziegler},
here we applied a Gaussian convolution with a fixed intensity and broadening to graphically compare the calculated and the ML-predicted vibrational frequencies.
As an example, Fig.\ref{fig:QM9_vibrfreq} shows a comparison between the harmonic frequencies calculated using B3LYP with the cc-pVDZ basis set and the ML predictions for molecule Nr 10,000 of QM9 ((1-cyano-2-oxoethyl)formate) . 
From the cross-validation scores, shown in section \ref{sec:CV}, it is no surprise that the frequencies associated with bond stretching are predicted fairly accurately by the machine learning method proposed. Around 3000 $cm^{-1}$ one clearly recognizes the C-H stretching modes, which differ from the calculated ones by less than 5$cm^{-1}$. The vibrational stretchings of the cyano group(2360 $cm^{-1}$) and the carbonyl groups(1840-1890 $cm^{-1}$), as well as the scissoring of the carbonyl O=C-H angles, are predicted with an error lower than 10$cm^{-1}$.
For lower frequency vibrations the predictions are not as accurate for mainly two reasons: the error in the prediction of Hessian elements associated with angles and dihedrals is larger than for bonds, and, more importantly, such vibrations are strongly non-local, and involve larger parts of the molecule. It is not easy to predict these vibrations using just a local representation.

Within the harmonic approximation, the sum of the vibrational frequencies is the vibrational zero point energy (ZPE) of the molecules $ZPE= (1/2) \hbar  \sum_i \omega_i $. This quantity represents the ground state vibrational energy of a molecule and can be easily computed from both the calculated and the ML predicted Hessian. In table \ref{tab:ZPEs} we summarize the results for the 13 QM9 molecules analyzed: the ML approximations always underestimate the ZPE with an error which ranges from 1.3\% to 14.5\%.
\begin{table}[h!]
\begin{tabular*}{.95\linewidth} {@{\extracolsep{\fill}}r|c|c|r}
QM9 Index   &   Calculated ZPE & Predicted ZPE & Error \\
\hline
10,000   &   68.2198  &  67.3312  &  1.3 \%  \\
20,000   &   93.9252  &  89.7762  &  4.42 \%  \\
30,000   &   126.745  &  123.158  &  2.83 \%  \\
40,000   &   208.222  &  178.019  &  14.5 \%  \\
50,000   &   122.161  &  108.193  &  11.4 \%  \\
60,000   &   203.131  &  186.773  &  8.05 \%  \\
70,000   &   160.905  &  139.64  &  13.2 \%  \\
80,000   &   170.515  &  162.541  &  4.68 \%  \\
90,000   &   157.959  &  144.444  &  8.56 \%  \\
100,000   &   154.283  &  149.215  &  3.28 \%  \\
110,000   &   167.916  &  155.576  &  7.35 \%  \\
120,000   &   169.058  &  157.315  &  6.95 \%  \\
130,000   &   92.2275  &  88.5746  &  3.96 \%  \\
\end{tabular*}
\label{tab:ZPEs}
\caption[width=\linewidth]{Vibrational ZPE for 13 QM9 molecules whose indices are multiples of 10,000 predicted by ML predictions and calculated explicit with B3LYP-DFT. Energy values are expressed in atomic units (milli Hartree ).}
\end {table}

\section{Conclusion}
We proposed a machine-learning based method to approximate diagonal as well as non-diagonal elements of the Hessian of a molecule. The representation used is specific for every internal coordinates, and takes explicitly into account the bond order, which is sensible because a single point DFT calculation is computationally considerably less expensive that the explicit calculation of the Hessian.
We trained our ML model on a relatively small dataset (subset of QM7) of less than 7000 molecules. The Hessian was computed at the B3LYP/cc-pVDZ level of theory. 
The agreement between ML and DFT was satisfactory. In particular, the calculated MAPE for bond stretching force constant was below 2\%, and was particularly small for bonds involving hydrogen atoms because they point outwards and are less affected by the chemical environment. The MAPE for bending and torsion was of 5\% and 10\%, respectively. 
From the ML model trained on QM7 we were also able to predict the Hessian of some molecules representative of the QM9 dataset. The Hessian predicted in internal coordinates was then transformed into the mass-weighted Cartesian Hessian, the diagonalization of which yields the harmonic vibrational frequencies and normal modes, that can be compared to the ones calculated  explicitly from DFT.

High frequency vibrations and normal modes were predicted accurately, while lower frequency ones were not. This behaviour is analogous to the IR spectroscopy theory, where stretchings and bendings can be identified accurately, while torsion and delocalized vibrations are more difficult to be interpreted.

The approximate Hessian obtained with ML is computational inexpensive and can be used as an initial Hessian guess for geometry optimization, or in the context of alchemical geometry relaxation \cite{Domenichini2020,domenichini2022alchemical, shiraogawa2022exploration,shiraogawa2023optimization}. 
A good starting Hessian may speed up the convergence of the geometrical optimization. An in detail assessment of the performance of the ML Hessian proposed is not yet provided, but should carefully take into account many parameters on which the optimization depends, \textit{e.g.} the type of molecule, the initial geometry, the optimization algorithm, and the Hessian update scheme.

\section{Acknowledgements}

We acknowledge financial support of the Austrian Science Fund (FWF) through the SFB TACO, Grant number F 81-N. The computational results presented have been achieved in part using the Vienna Scientific Cluster (VSC).

We also would like to thank Jan Weinreich, Michail Sahre, and Luigi Ranalli, from the University of Vienna, as well as Danish Khan and Anatole von Lilienfeld from the University of Toronto, for the productive discussions and feedback received.

\section{References}
\bibliography{bib_files/BasisRef,bib_files/Alchemy_vLg,bib_files/Alchemy_Others, bib_files/Alchemy_Cardenas, bib_files/Alchemy_Geerlings, bib_files/Alchemy_Keith, bib_files/software, bib_files/Other_works ,bib_files/QML, bib_files/derivatives,bib_files/Geomopt,bib_files/Dellago,bib_files/behler_parrinello,bib_files/randomforest,bib_files/IRspectroscopy,bib_files/Geometryoptimizations_review,bib_files/schlegel,bib_files/Barone_ML,bib_files/Heiniswork,bib_files/MLhessianspectra,bib_files/Ceotto_ML,bib_files/Domenichini,bib_files/Alchemy_Hasegawa, bib_files/datasets}

\end{document}


\title{Supplementary Information for: Molecular Hessian matrices from a machine learning random forest regression algorithm}

\author{Giorgio Domenichini} 
\affiliation{Faculty of Physics,
University of Vienna,
Kolingasse 14-16,
1090 Vienna,
Austria}

\author{Christoph Dellago} 
\affiliation{Faculty of Physics,
University of Vienna,
Kolingasse 14-16,
1090 Vienna,
Austria}

\maketitle
\section{Supplementary tables}

The following tables show the mean average percentage error, averaged over ten train test split of the QM7 DataSet, for the Hessian's terms associated to various types of bonds stretchings, angle bendings, and dihedral torsions, the coordinates were classified according to their elements type and if or if not are part (with two atoms or more ) of an atomic ring system. For every group are enlisted its occurrence in the dataset as well the resulting MAPE. It was impossible to train a random forest on groups with less than 10 members.    
\begin{table}
  
\begin{tabular}{l|c|c|c|c}
\toprule
{} &        Nr.Acycl. &    MAPE Acycl.& Nr.Cycl.  &MAPE Cycl.\\
\midrule
S-N &     14 &  5.241632 &     86 &  4.631084 \\
S-C &    118 &  3.389312 &    333 &  2.131339 \\
O-N &    385 &  1.729843 &    219 &  4.760662 \\
N-N &    313 &  1.457288 &    220 &  3.906045 \\
S-O &    188 &  1.018311 &      5 &  7.433617 \\
N-C &   5315 &  0.680587 &   4137 &  2.727756 \\
C-C &  14441 &  0.634709 &  11005 &  2.176796 \\
O-C &   4681 &   0.57737 &   2498 &  2.699537 \\
O-H &     1679 &  0.181918 &     0 &           \\
N-H &     5265 &   0.17556 &     0 &           \\
C-H &    52203 &  0.140281 &     0 &           \\
\bottomrule
\end{tabular}

\end{table}

\begin{table}
\begin{tabular}{l|c|c|c|c}
\toprule
{} &        Nr.Acycl. &    MAPE Acycl.& Nr.Cycl.  &MAPE Cycl.\\
\midrule
S-C-C &     60 &  28.166632 &    344 &    4.45253 \\
S-C-O &      9 &   26.48468 &     28 &  39.773029 \\
S-N-C &      7 &  21.458166 &     85 &   5.746084 \\
S-C-N &     34 &  21.094845 &    114 &   9.303313 \\
O-S-N &     28 &   21.01224 &     14 &  11.006566 \\
O-S-C &    204 &  16.973637 &     89 &  12.073264 \\
O-S-O &    112 &  16.004671 &     10 &  29.536047 \\
C-S-C &     35 &  12.001598 &    126 &   2.353975 \\
O-C-O &    164 &   9.444957 &    167 &   6.422588 \\
C-C-C &   8985 &   8.967821 &  13821 &   4.510938 \\
N-S-C &     14 &   8.046433 &     76 &   2.199812 \\
N-N-C &    509 &   6.829266 &    554 &   5.818407 \\
C-N-C &   1191 &   5.808174 &   2486 &   5.865337 \\
S-O-C &      5 &   5.213844 &      5 &  25.685764 \\
N-C-C &   3700 &   5.149424 &   6156 &   4.997315 \\
S-N-H &     19 &   4.995772 &     12 &  18.130427 \\
O-C-C &   3677 &   4.495432 &   4910 &   4.985185 \\
S-C-H &    158 &   4.477941 &    238 &   4.471362 \\
O-N-C &    390 &   3.751602 &    282 &   5.408519 \\
N-C-N &    360 &   3.731582 &    289 &   4.010244 \\
N-O-C &    142 &   3.580214 &    209 &   4.501361 \\
C-O-C &    737 &   3.002897 &   1142 &   3.987935 \\
O-C-N &    485 &     2.3672 &    410 &   2.518777 \\
O-N-H &     41 &    2.17109 &     22 &   7.827331 \\
N-N-H &    358 &    1.82311 &    121 &   4.000126 \\
C-C-H &  38571 &     1.7042 &  27504 &   2.198234 \\
C-N-H &   4556 &   1.505966 &   2011 &   3.124657 \\
N-C-H &   6528 &   1.461551 &   4806 &   2.583123 \\
O-C-H &   5299 &   1.309337 &   3168 &   2.385833 \\
S-O-H &      7 &   0.978599 &      0 &            \\
C-O-H &     1431 &   0.878634      0 &            \\
H-C-H &    29776 &   0.815284 &    0 &            \\
N-O-H &      241 &    0.77457 &    0 &            \\
H-N-H &     1531 &   0.717418 &    0 &            \\
N-O-N &        0 &            &    5 &   0.990445 \\
O-N-N &        0 &            &    11 &   6.098738 \\
N-S-N &        0 &            &     5 &   0.980887 \\
S-N-N &        0 &            &     7 &   0.544051 \\
\bottomrule
\end{tabular}

\end{table}

\begin{table}
\begin{tabular}{l|c|c|c|c}
\toprule
{} &        Nr.Acycl. &    MAPE Acycl.& Nr.Cycl.  &MAPE Cycl.\\
\midrule
C-C-C-H &  27348 &    3.747194 &  40645 &    3.034742 \\
H-C-C-H &  22100 &    1.669997 &  16983 &    1.904789 \\
O-C-C-H &   5467 &    4.110079 &   6807 &    4.085248 \\
H-N-C-H &   5313 &    2.540041 &   2430 &    3.256849 \\
N-C-C-H &   5115 &    4.498609 &   8574 &    4.089736 \\
C-N-C-H &   4274 &    4.647676 &   6954 &    4.340487 \\
H-N-C-C &   3624 &     7.78624 &   4636 &    7.001287 \\
C-C-C-C &   3329 &    8.945028 &  10336 &     5.73141 \\
C-O-C-H &   2535 &    3.991645 &   2902 &    3.644295 \\
O-C-C-C &   2199 &    8.062476 &   5171 &    6.816527 \\
N-C-C-C &   1754 &   11.083956 &   5797 &    6.913438 \\
H-O-C-H &   1744 &    2.481936 &      0 &             \\
H-O-C-C &   1387 &    7.937584 &    910 &    7.834636 \\
C-N-C-C &   1211 &   14.215368 &   4698 &    8.441531 \\
C-O-C-C &   1004 &   16.451115 &   2700 &    9.918548 \\
N-N-C-H &    725 &    9.093003 &    623 &   11.679359 \\
H-N-C-N &    634 &   10.217266 &    313 &   13.121612 \\
O-C-C-N &    433 &   10.663836 &    924 &   10.649038 \\
H-N-C-O &    432 &    11.09593 &    224 &   14.576883 \\
C-N-N-H &    358 &    8.883018 &    174 &   24.431662 \\
O-N-C-H &    355 &    8.869474 &    213 &   17.089795 \\
N-N-C-C &    336 &   24.538703 &    666 &    17.43391 \\
O-S-C-H &    334 &    3.691449 &    148 &    4.789884 \\
C-N-C-N &    328 &   18.529809 &    580 &   18.974457 \\
O-N-C-C &    309 &   13.957791 &    316 &   17.214357 \\
N-O-C-H &    306 &    6.864002 &    258 &    9.503669 \\
C-N-C-O &    279 &   17.084418 &    417 &   16.799575 \\
H-O-N-C &    242 &    9.095877 &     51 &   16.557715 \\
C-N-N-C &    193 &   35.602058 &    318 &   19.521871 \\
O-C-C-O &    188 &   11.240148 &    348 &   13.875503 \\
\bottomrule
\end{tabular}
\end{table}

\begin{table}
\begin{tabular}{l|c|c|c|c}
\toprule
{} &        Nr.Acycl. &    MAPE Acycl.& Nr.Cycl.  &MAPE Cycl.\\
\midrule
C-O-N-C &    146 &   23.325096 &    221 &   31.548462 \\
C-O-C-O &    126 &   24.838712 &    225 &   18.365464 \\
C-S-C-H &    116 &     8.05075 &    181 &    6.534324 \\
O-S-C-C &    110 &    5.198317 &    100 &    7.264619 \\
N-C-C-N &    105 &    7.592818 &    399 &   14.234377 \\
S-C-C-H &    105 &    5.073537 &    363 &     7.19933 \\
N-O-C-C &     80 &   25.196806 &    276 &   18.636077 \\
N-N-C-O &     73 &    20.58332 &     60 &   22.165334 \\
C-O-C-N &     73 &   25.375109 &    174 &   21.034904 \\
O-N-C-N &     61 &    10.67592 &     51 &   36.792841 \\
H-O-C-O &     52 &   14.163598 &     11 &   16.962884 \\
O-N-C-O &     47 &   15.977449 &     40 &   39.396365 \\
O-S-N-H &       38 &    2.760768 &       10 &             \\
H-N-C-S &     36 &   23.522447 &     61 &   11.645979 \\
C-S-C-C &     30 &    11.82768 &    242 &   12.842329 \\
N-S-C-H &     24 &    7.074261 &     49 &   11.661814 \\
H-N-N-H &       23 &    9.137747 &        8 &             \\
H-O-N-H &       22 &    5.169125 &        0 &             \\
C-N-C-S &     20 &   48.093787 &     98 &    17.34324 \\
C-S-N-H &     19 &    9.834999 &     12 &   97.659343 \\
C-O-N-H &     19 &   10.260201 &     22 &   81.359734 \\
O-S-C-O &       18 &    6.007688 &       10 &             \\
S-N-C-H &     18 &   20.778762 &     61 &   14.595413 \\
O-S-N-C &     14 &    5.106503 &     20 &   15.509021 \\
O-S-O-H &       14 &    5.001918 &        0 &             \\
N-O-C-O &     12 &  119.676177 &     27 &   46.213247 \\
S-O-C-H &       12 &   21.719571 &        8 &             \\
N-S-C-C &       10 &             &       94 &   13.101342 \\
S-C-C-C &        9 &             &      271 &   10.064542 \\
C-S-C-O &        9 &             &       22 &   34.561446 \\
C-S-N-C &        7 &             &       75 &   13.159854 \\
H-O-C-S &        4 &             &       12 &   28.698307 \\
S-C-C-O &        2 &             &       30 &   24.456751 \\
O-S-C-N &        2 &             &       14 &   10.519502 \\
S-N-C-C &        2 &             &      104 &   16.806332 \\
C-S-C-N &        1 &             &       90 &   10.883112 \\
S-C-C-N &        1 &             &       90 &   11.916005 \\
O-N-N-C &        0 &             &       11 &   37.010733 \\
N-O-C-N &        0 &             &       11 &     40.6053 \\
S-N-C-N &        0 &             &       12 &   31.456192 \\
H-O-C-N &        0 &             &       28 &  155.818809 \\
\bottomrule
\end{tabular}
\end{table}

\section{Supplementary figures}

\begin{figure}
    \centering
   \includegraphics[width=.48\linewidth]{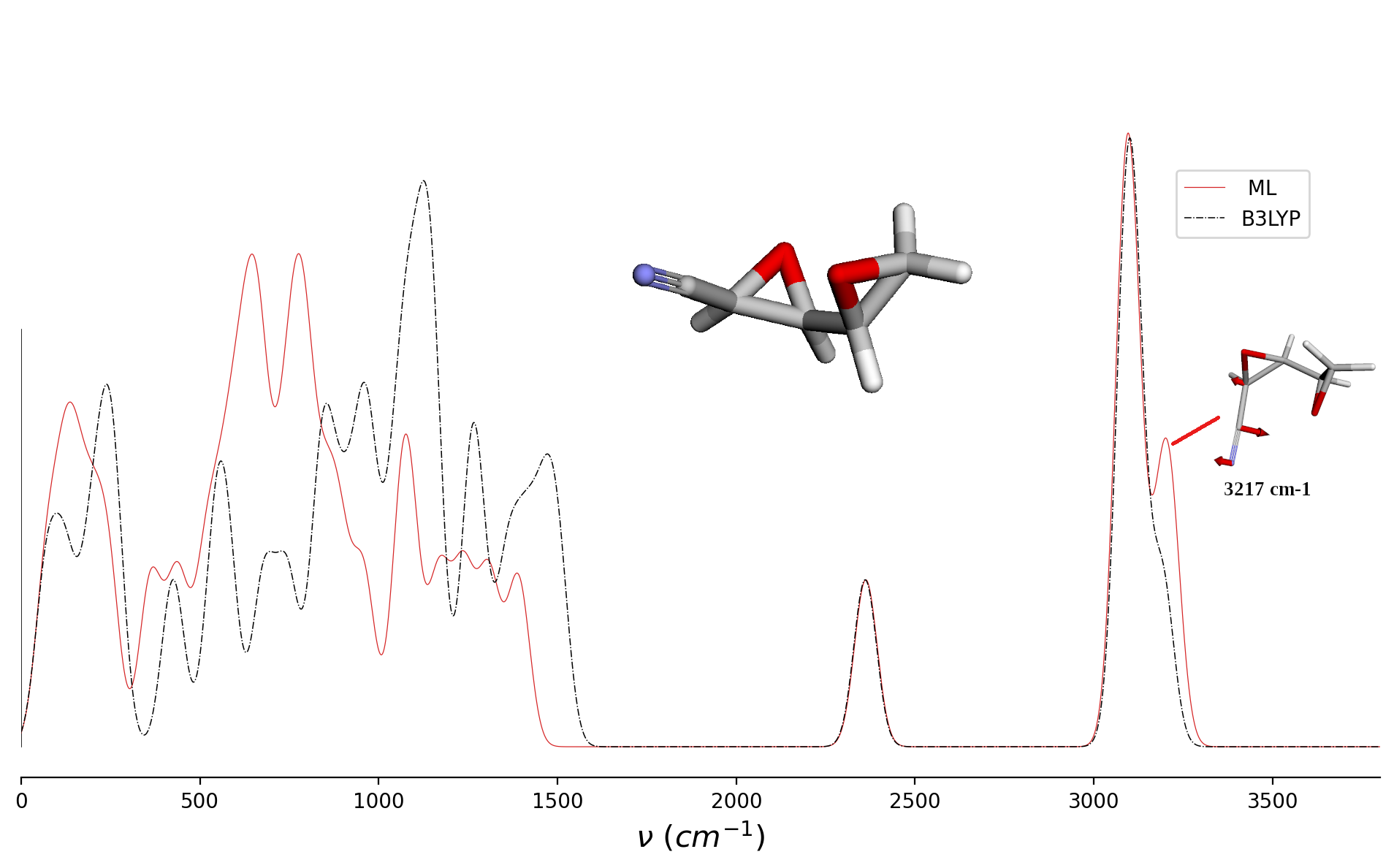}
   \includegraphics[width=.48\linewidth]{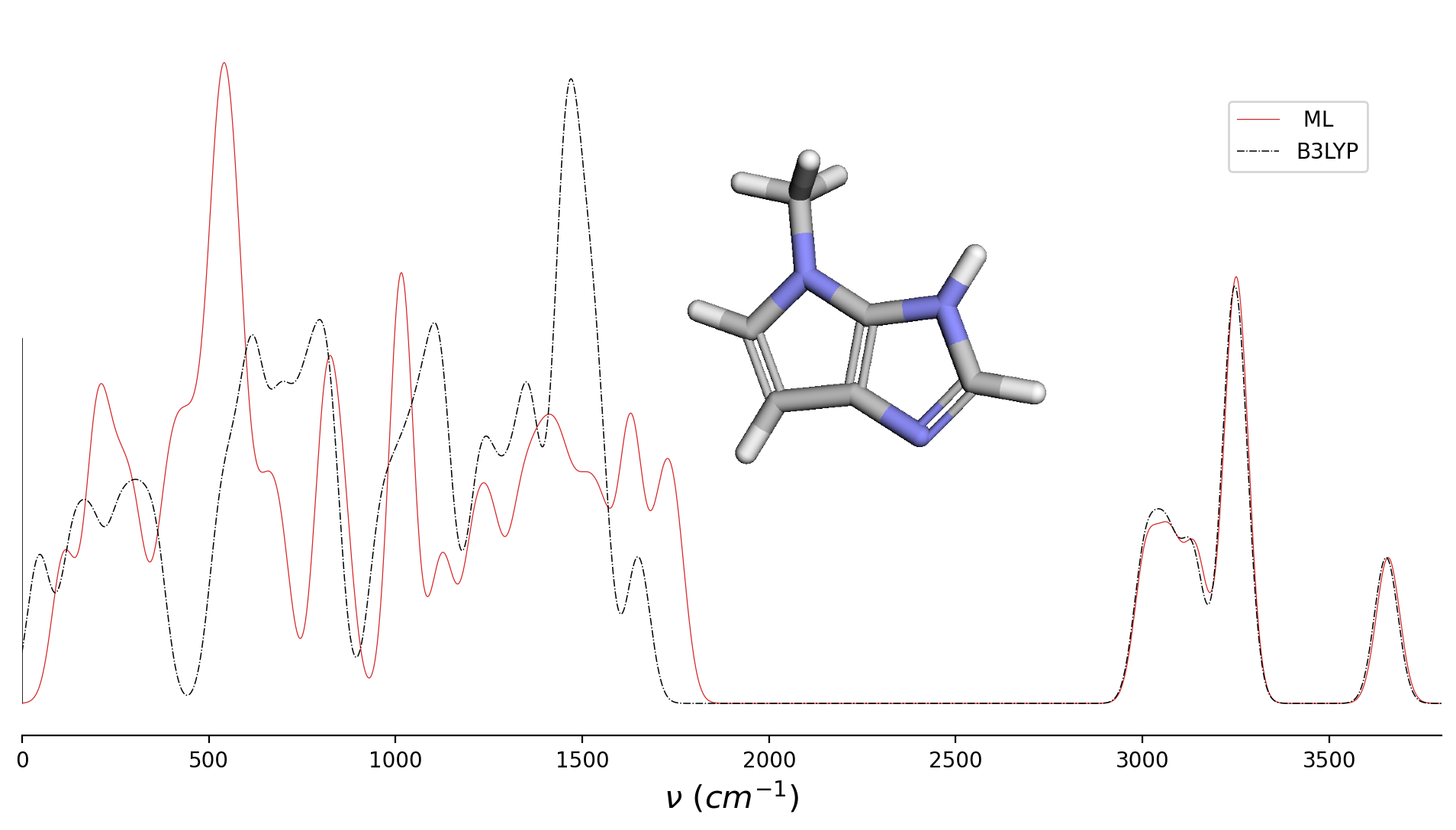}
   \includegraphics[width=.48\linewidth]{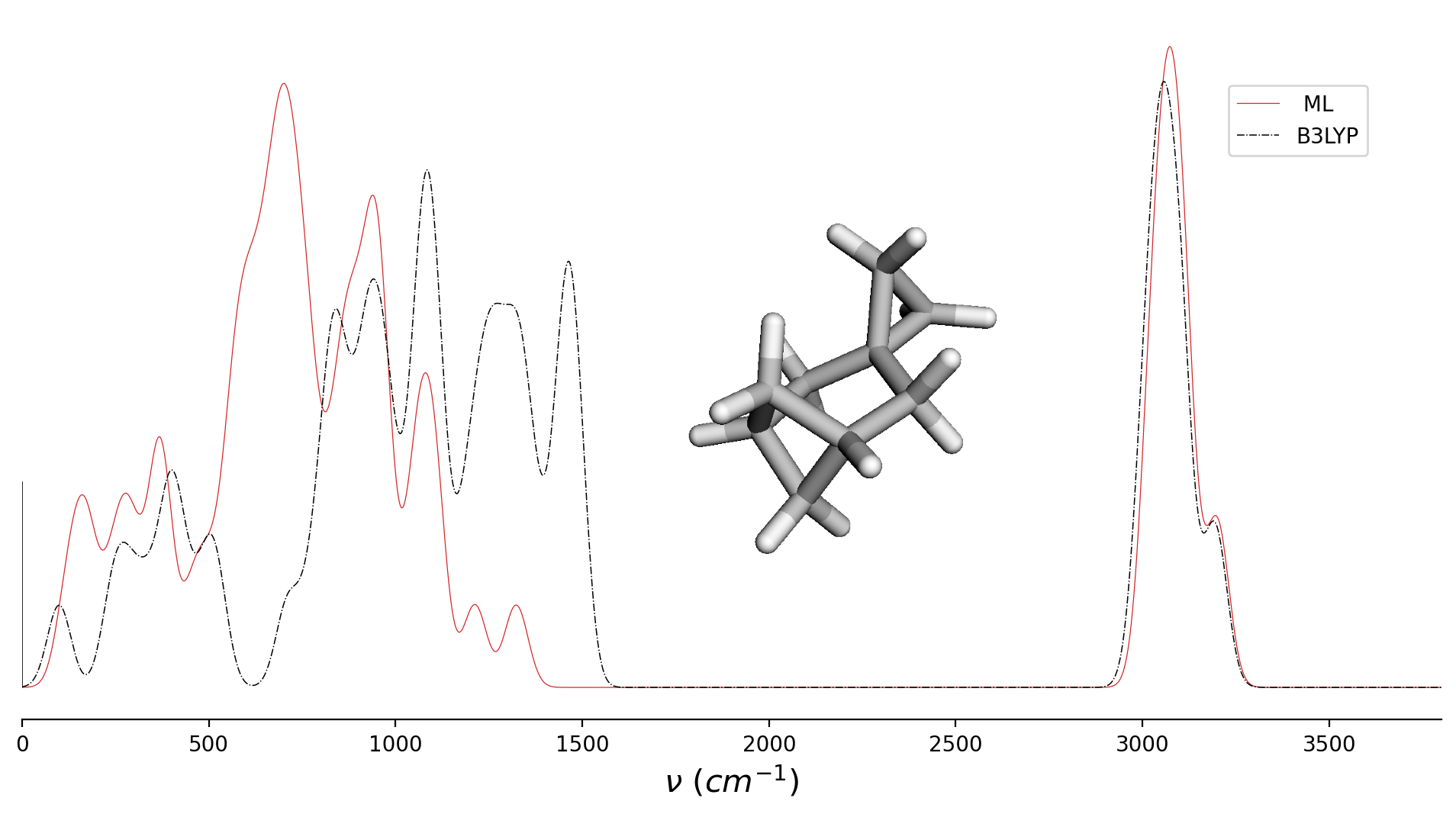}
   \includegraphics[width=.48\linewidth]{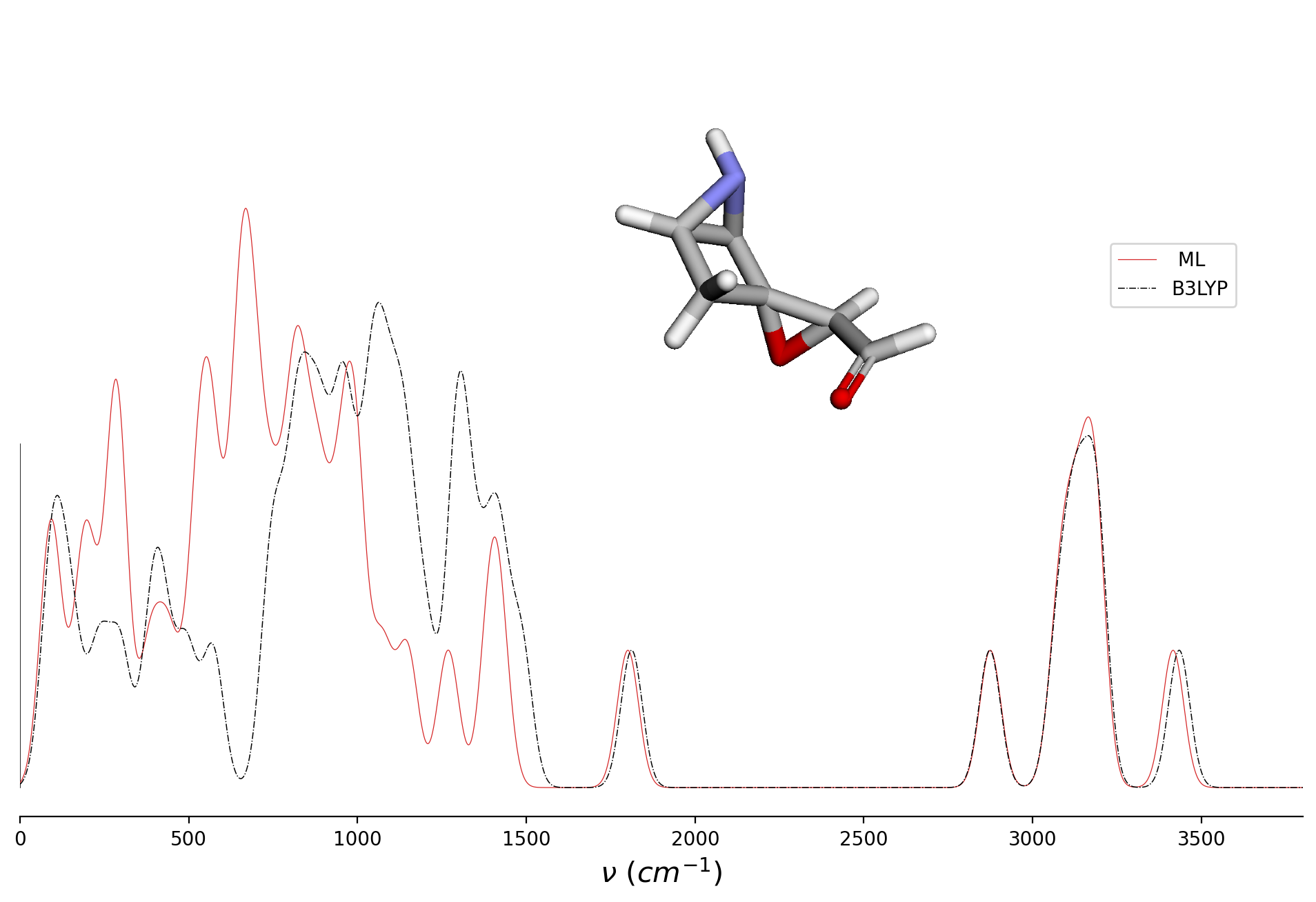}
   \includegraphics[width=.48\linewidth]{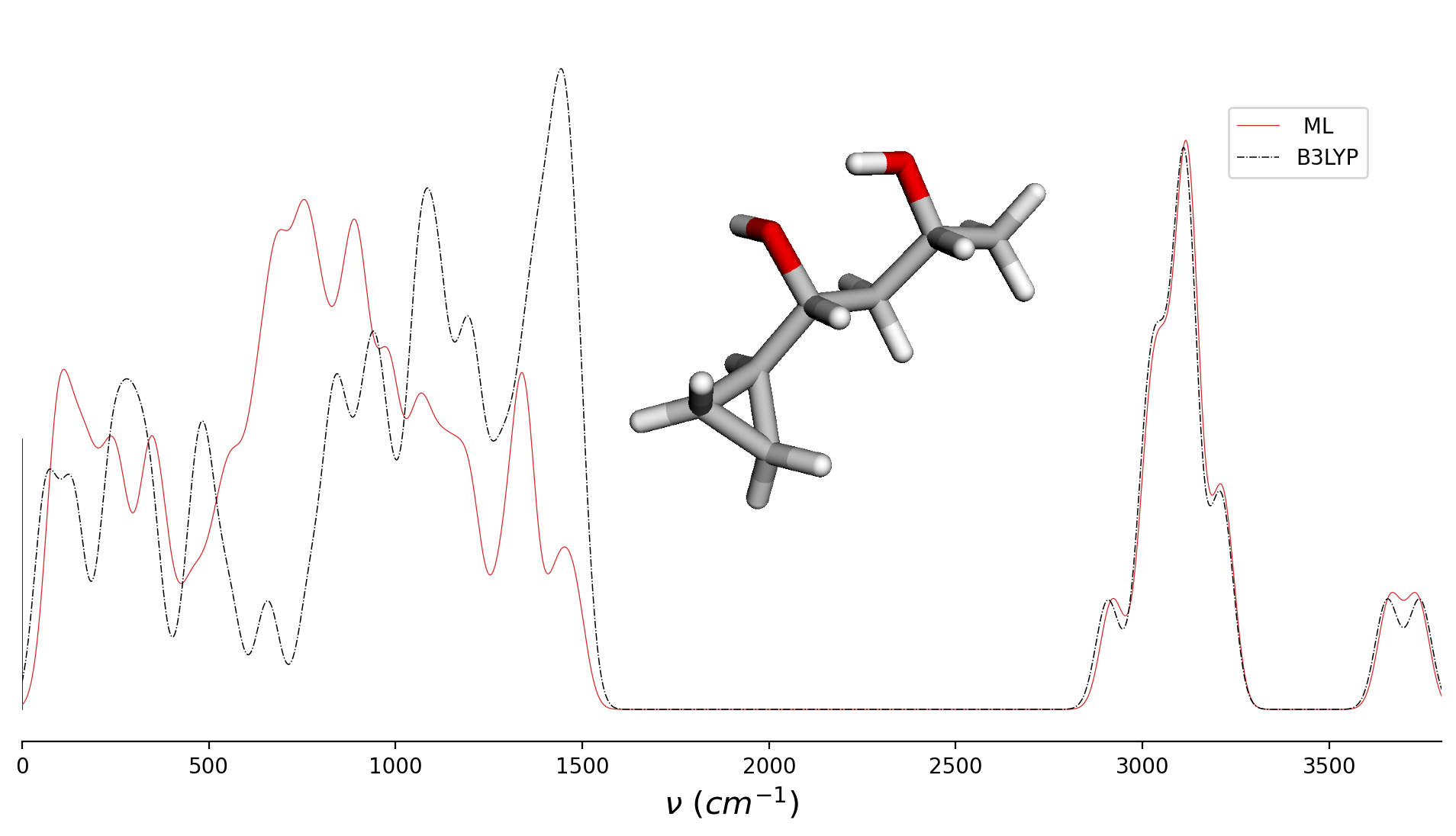}
   \includegraphics[width=.48\linewidth]{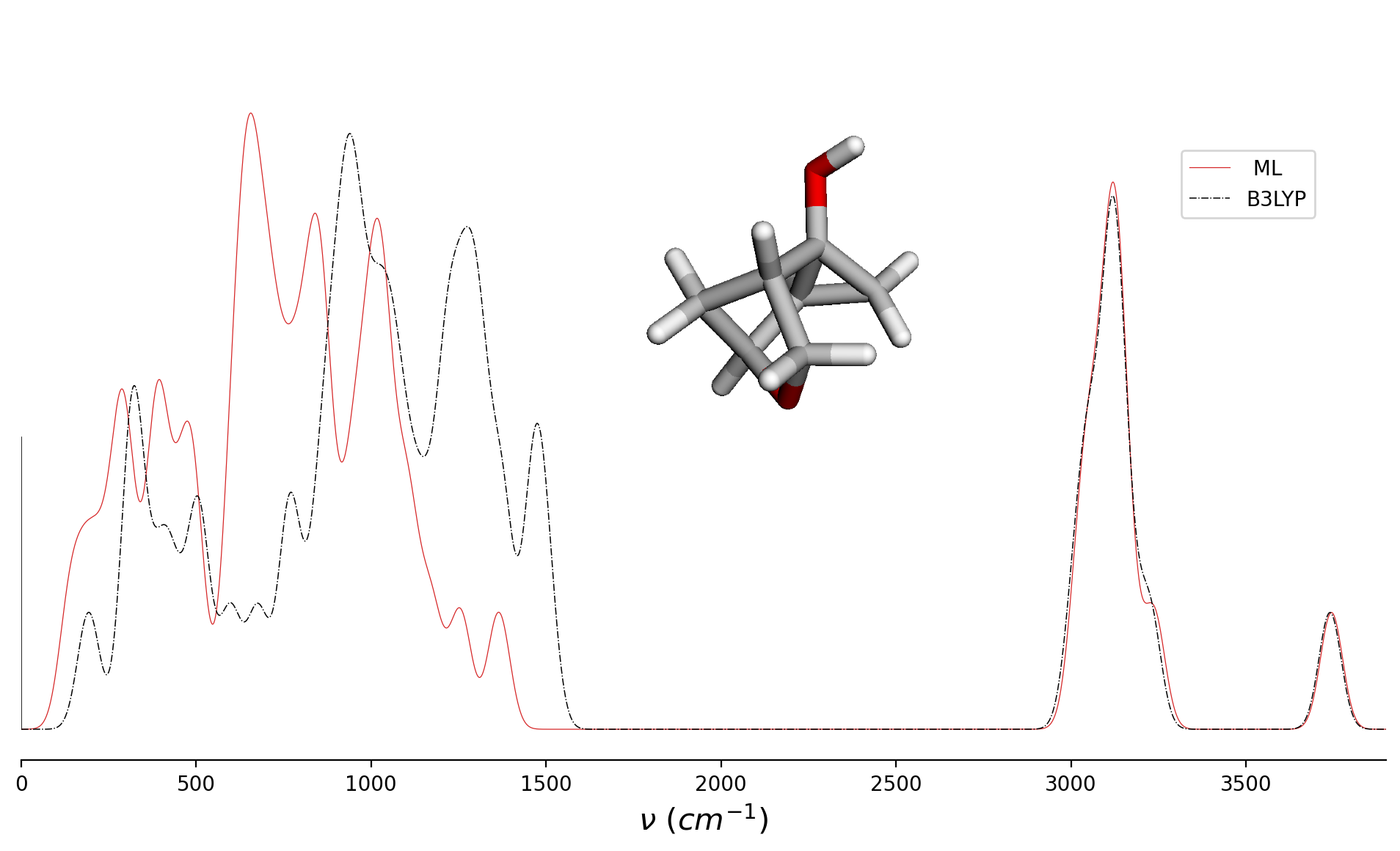}
   \includegraphics[width=.48\linewidth]{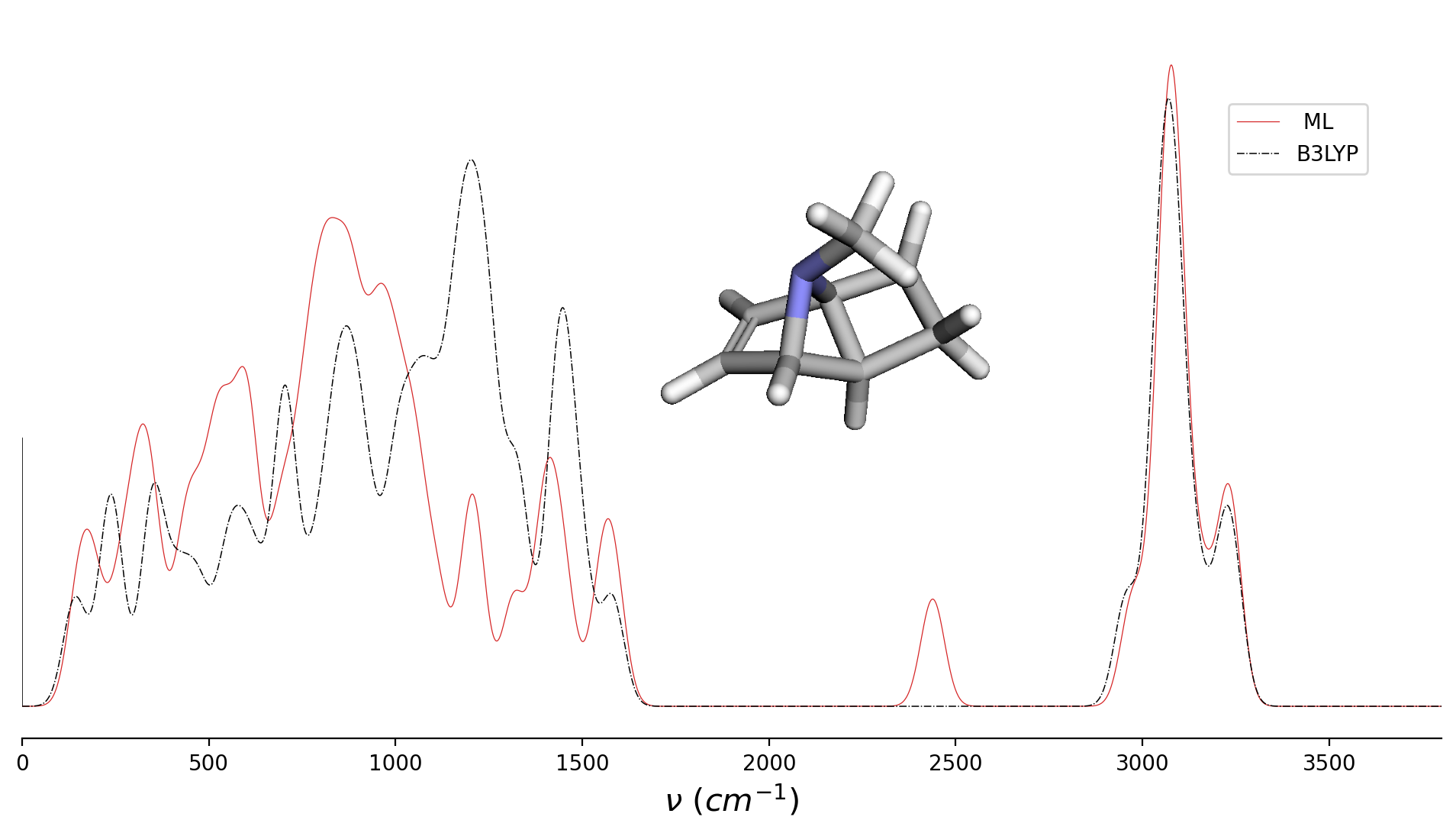}
   \includegraphics[width=.48\linewidth]{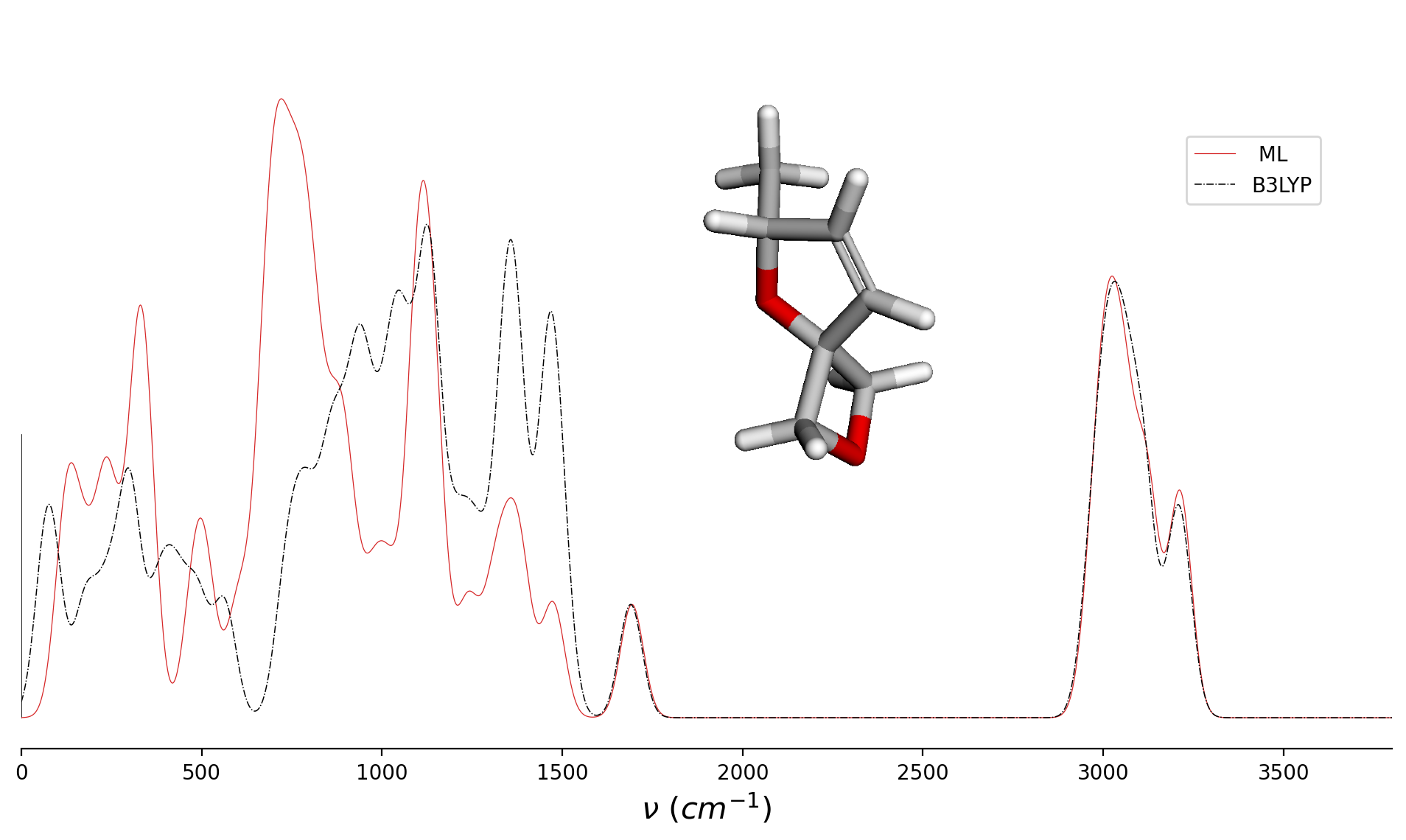}
   \includegraphics[width=.48\linewidth]{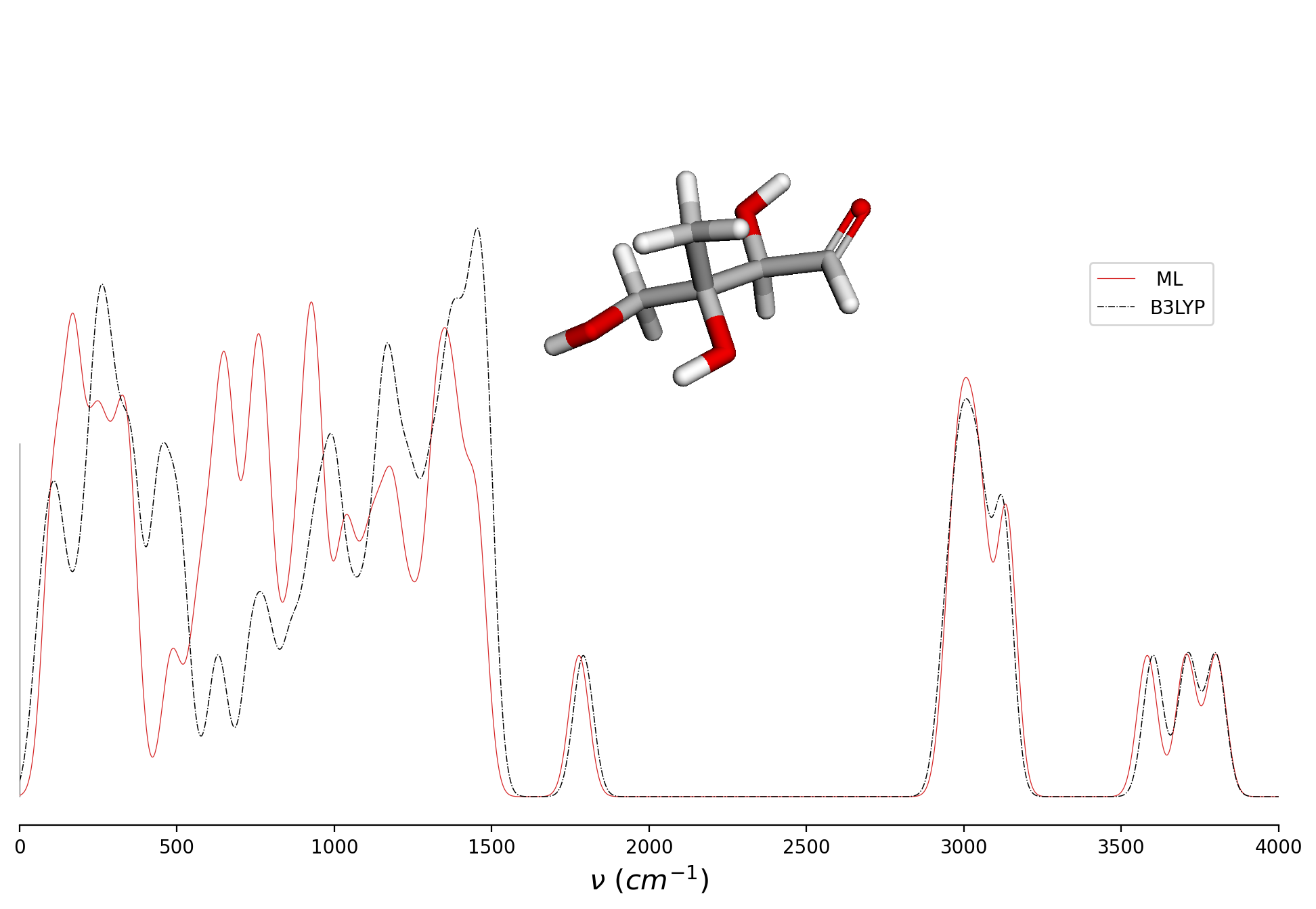}
   \includegraphics[width=.48\linewidth]{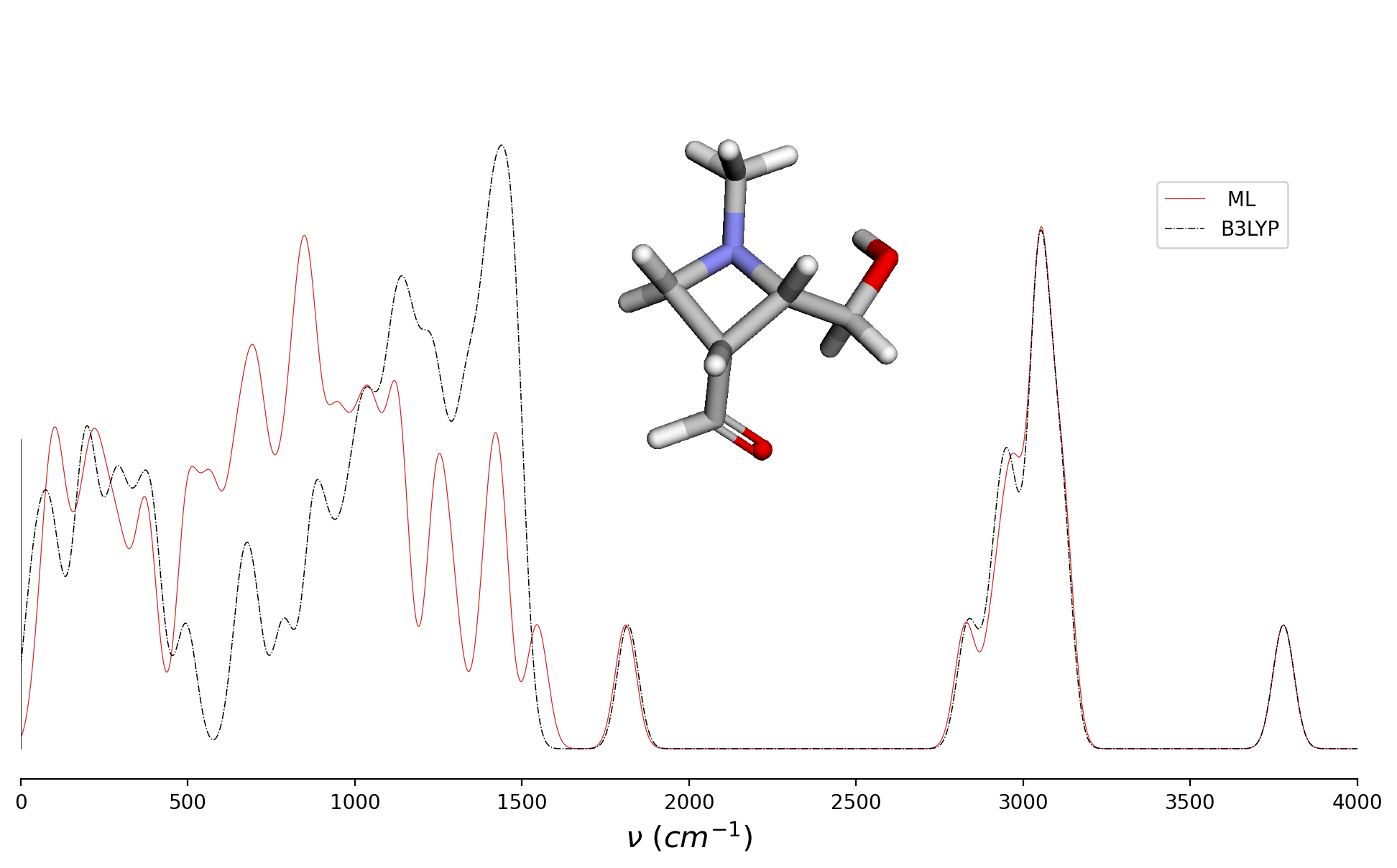}
   \includegraphics[width=.48\linewidth]{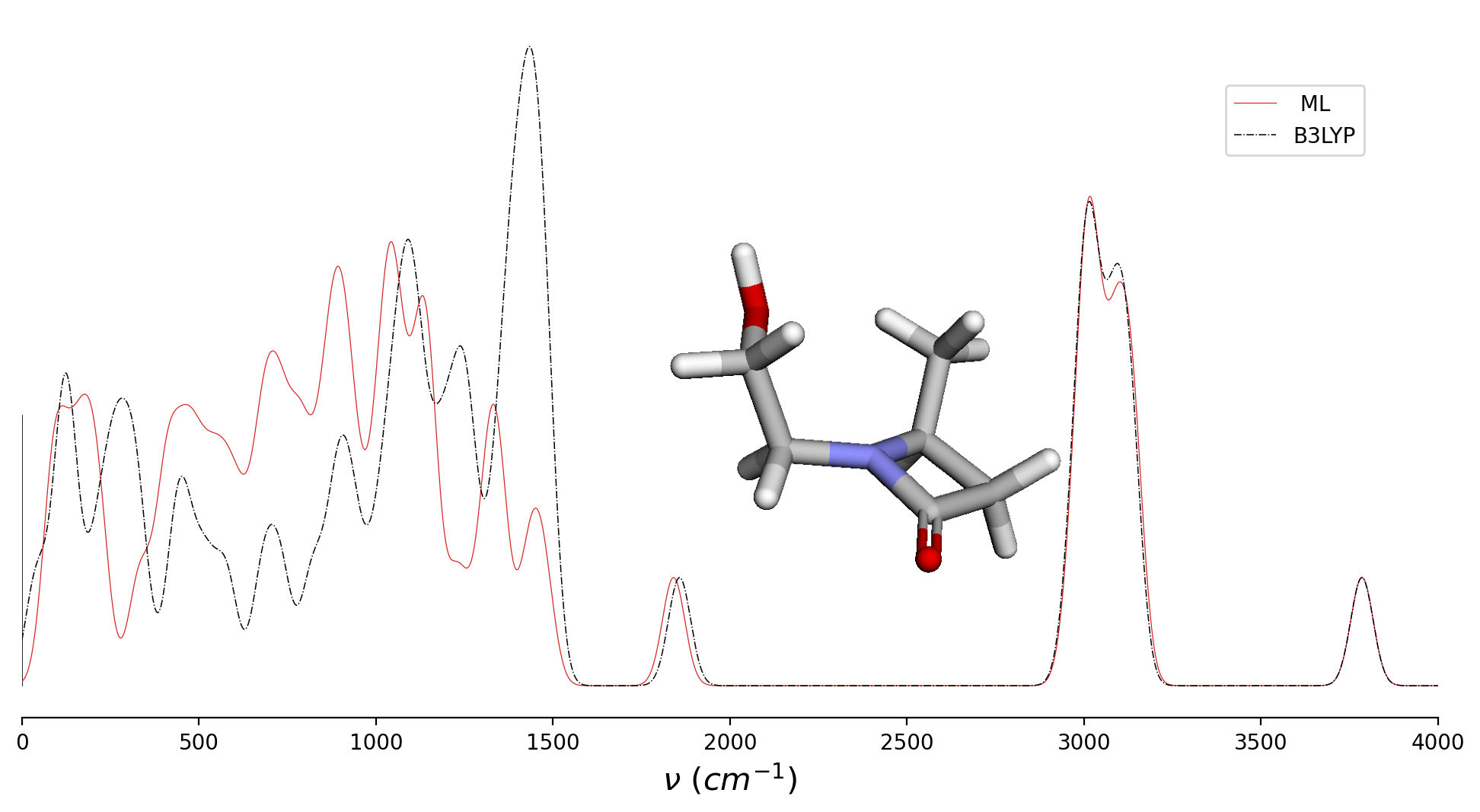}
   \includegraphics[width=.48\linewidth]{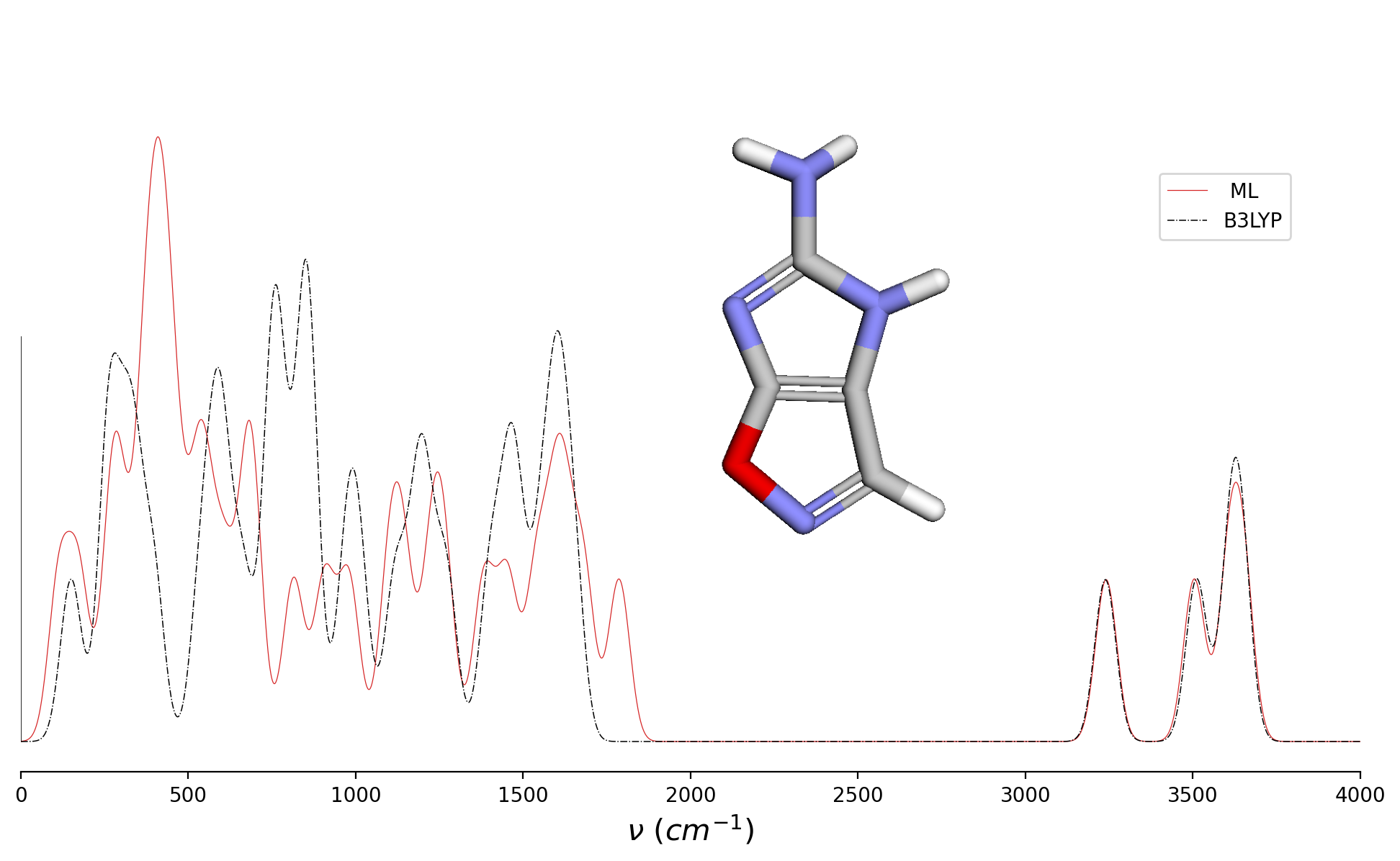}
    \caption{Predicted vibrational frequencies for the QM9 molecules, whose index it is a multiple of 10000. For molecule 20000 in the top left panel there is a high frequency outlier. The error seems to be related to the N-C-C angle bending, but in reality it is due to the wrong ML predictions of a diheadral torsion and to the omission of the dihedral-dihedral non-diagonal terms in the ML prediction. The near-linearity of the N-C-C angle amplifies the error. }
    \label{fig:my_label}  
\end{figure}

\begin{figure} [h!]
    \centering
    \includegraphics[width=\linewidth]{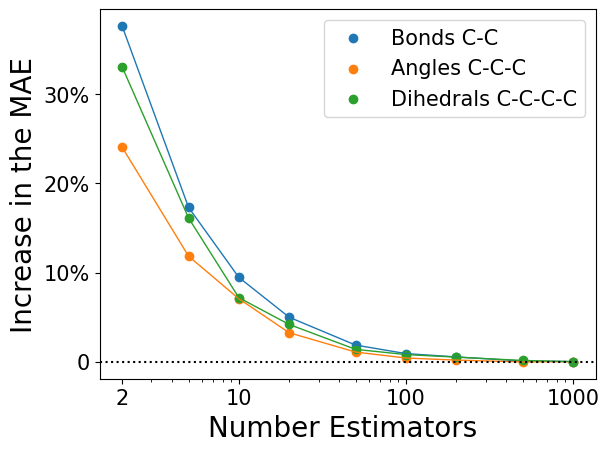}
    \caption{Dependence of the Mean Absolute Error (MAE) of the Hessian for bonds, angle and dihedrals involving only carbon atoms on the number of estimators in the random forest. Results are averaged over 10 train test splits of the QM7 dataset. The baseline is the MAE for 1000 estimators. The error for 100 estimators is just 1\% higher than the error for 1000, and therefore a good choice between precision and computing time.}
    \label{fig:N_Extimators}
\end{figure}

%


\bibliography{bib_files/BasisRef,bib_files/Alchemy_vLg,bib_files/Alchemy_Others,bib_files/software,bib_files/Other_works,bib_files/QML,bib_files/derivatives,bib_files/Geomopt}